\documentclass[prd,preprint,nofootinbib]{revtex4-1}

\usepackage{color}
\usepackage[latin1]{inputenc}
\usepackage[T1]{fontenc}
\usepackage{amsmath}
\usepackage{amssymb}
\usepackage{amsthm}
\usepackage{mathrsfs}
\usepackage{fancyhdr}
\usepackage{graphicx}
\usepackage{epstopdf}
\usepackage{subfigure}
\usepackage{placeins}
\usepackage{array}
\usepackage[all]{xy}
\usepackage{eufrak}
\usepackage{euscript}
\usepackage{enumerate}
\usepackage{slashed}
\usepackage{hyperref}
\usepackage{hypcap}

\hypersetup{pdftex,colorlinks=true,linkcolor=blue,citecolor=blue,menucolor=black,urlcolor=blue,filecolor=blue}

\begin{document}

\title{Thermal photon and dilepton production and electric charge transport in a baryon rich strongly coupled QGP from holography}

\author{Stefano Ivo Finazzo}
\email{stefano@ift.unesp.br}
\affiliation{Instituto de F\'{i}sica Te\'orica, Universidade do Estado de S\~{a}o Paulo, Rua Dr. Bento T. Ferraz 271, 01140-070, S\~{a}o Paulo, SP, Brazil}

\author{Romulo Rougemont}
\email{romulo@if.usp.br}
\affiliation{Instituto de F\'{i}sica, Universidade de S\~{a}o Paulo, C.P. 66318, 05315-970, S\~{a}o Paulo, SP, Brazil}


\begin{abstract}
We obtain the thermal photon and dilepton production rates in a strongly coupled quark-gluon plasma (QGP) at both zero and nonzero baryon chemical potential using a bottom-up Einstein-Maxwell-Dilaton (EMD) holographic model that is in good quantitative agreement with the thermodynamics of $(2+1)$-flavor lattice QCD around the crossover transition for baryon chemical potentials up to 400 MeV, which may be reached in the beam energy scan (BES) at RHIC. We find that increasing the temperature $T$ and the baryon chemical potential $\mu_B$ enhances the peak present in both spectra. We also obtain the electric charge susceptibility, the DC and AC electric conductivities and the electric charge diffusion as functions of $T$ and $\mu_B$. We find that electric diffusive transport is suppressed as one increases $\mu_B$. At zero baryon density, we compare our results for the DC electric conductivity and the electric charge diffusion with the latest lattice data available for these observables and find reasonable agreement around the crossover transition. Therefore, our holographic results may be used to constraint the magnitude of the thermal photon and dilepton production rates in a strongly coupled QGP, which we found to be at least one order of magnitude below perturbative estimates.
\end{abstract}

\maketitle
\tableofcontents

\section{Introduction}

Some of the cleanest observables in ultrarelativistic heavy ion collisions which provide sensitive information about the initial state conditions of the quark-gluon plasma (QGP) \cite{Adcox:2004mh,Arsene:2004fa,Back:2004je,Adams:2005dq,Aad:2013xma,Gyulassy:2004zy,Heinz:2013th,Shuryak:2014zxa} are the spectra of photons and dileptons. The fact that photons and leptons do not couple via strong interactions to the quarks and gluons of the QGP implies that they escape away from the medium after weakly interacting with the bulk of the plasma, carrying local information about it \cite{Shuryak:1978ij}. These probes can thus be used to further constrain estimates for QGP transport coefficients, which explains, for instance, the recent interest on their measurements at RHIC (Relativistic Heavy Ion Collider) \cite{Adare:2008ab,Adare:2011zr,Adare:2014fwh} and LHC (Large Hadron Collider) \cite{Wilde:2012wc,Lohner:2012ct}. One of the contributions to the total yield of these probes is from the thermal production of soft photons and dileptons in the QGP, mostly relevant at low transverse momentum $p_T \lesssim 3 \, \mathrm{GeV/c}$ \cite{Baier:1991em,Kapusta:1991qp}. At the highest energies currently reached in heavy ion collisions at RHIC and LHC, $\sqrt{s} = 200 \, \mathrm{GeV}$ and $\sqrt{s} = 2.76 \, \mathrm{TeV}$ per nucleon, respectively, the baryon chemical potential of the QGP is considerably smaller than its temperature ($\mu_B/T \ll 1 $). However, in order to probe the phase diagram of QCD in the $(T,\mu_B)$-plane and search for the critical end point of a putative line of first order chiral phase transition at nonzero values of $\mu_B$ \cite{Stephanov:1999zu,Stephanov:1998dy}, RHIC is performing a beam energy scan (BES) which goes down to energies between $\sqrt{s} = 7.7 \, \mathrm{GeV}$ and $\sqrt{s} = 62.4 \, \mathrm{GeV}$ per nucleon, prompting the need of computing the photon and dilepton production at finite temperature and baryon density.

Since the QGP produced at RHIC and LHC is expected to be strongly coupled around the QCD crossover transition \cite{Aoki:2006we}, non-perturbative methods are required in order to investigate the production of these electromagnetic probes in the plasma for temperatures in the range $T\sim 150\,\textrm{MeV} - 300\,\textrm{MeV}$. The use of lattice gauge theory methods to compute the needed production rates is also challenging, since one has to face the double challenge of dealing with the sign problem of the fermion determinant that arises at finite chemical potential and also the computation of real time correlation functions.\footnote{Albeit both challenges are currently being dealt with in some cases. See \cite{Aarts:2011ax,Fodor:2001au,Gavai:2003mf,Allton:2003vx,deForcrand:2003hx,deForcrand:2002ci,D'Elia:2002gd,Fodor:2009ax,Philipsen:2012nu} for some lattice strategies to circumvent the sign problem. See also \cite{Asakawa:2000tr} for a discussion of the Maximum Entropy Method (MEM) used to reconstruct spectral functions from lattice Euclidean correlators.} A useful non-perturbative tool to explore the strong coupling properties of non-Abelian gauge theories is the gauge/gravity duality \cite{Maldacena:1997re,Witten:1998qj,Witten:1998zw,Gubser:1998bc}, which may be naturally adapted in order to compute real time observables \cite{Son:2002sd,Herzog:2002pc,Gubser:2008sz,Skenderis:2008dg}. Since the thermal spectra of photons and dileptons requires the computation of the retarded propagator of the electric current, the gauge/gravity duality presents itself as a useful approach to study such phenomena.

The main objective of the present work is to present a holographic calculation of the thermal spectra of soft photons and dileptons in a strongly coupled non-conformal QGP, at both zero and nonzero $\mu_B$. This shall be done by using a bottom-up Einstein-Maxwell-Dilaton (EMD) holographic model \cite{Rougemont:2015wca} which provides a good quantitative description of the thermodynamics of $(2+1)$-flavor lattice QCD around the crossover transition for baryon chemical potentials up to 400 MeV (which is the maximum value of $\mu_B$ reached in the current BES at RHIC). This model lies within the class of bottom-up phenomenological EMD actions at finite chemical potential originally proposed in \cite{DeWolfe:2010he,DeWolfe:2011ts}, which in turn generalize the Eintein-Dilaton models at zero chemical potential proposed in \cite{Gubser:2008sz,Gubser:2008ny,Gubser:2008yx,Noronha:2009ud}.\footnote{For similar bottom-up constructions in the context of Improved Holographic QCD, see Ref.'s \cite{Gursoy:2007cb,Gursoy:2007er,Gursoy:2008bu,Gursoy:2008za,Gursoy:2009jd}. See also \cite{Charmousis:2010zz, Gouteraux:2011ce} for discussions of Effective Holographic Theories used in applications to condensed matter systems.} The thermal spectra we are going to compute here can then be used as inputs in calculations of the total photon and dilepton spectra, such as the ones pursued in Ref.'s \cite{Dion:2011pp,Chatterjee:2013naa,Shen:2013cca,Paquet:2015lta}. We shall also examine electric charge transport phenomena in the QGP at zero and nonzero $\mu_B$, extending the results of Ref. \cite{Finazzo:2013efa} to the $(T,\mu_B)$-plane.
 
In QCD, the thermal production rates have been computed at leading order in perturbation theory in Ref.'s \cite{Arnold:2001ba,Arnold:2001ms}, and in next-to-leading order in Ref. \cite{Ghiglieri:2013gia}. More recently, these thermal spectra have been also calculated in the context of the semi-QGP model \cite{Gale:2014dfa,Hidaka:2015ima}. Calculations of the thermal photon and dilepton spectra in $\mathcal{N} = 4$ Super-Yang Mills (SYM) were carried out in \cite{CaronHuot:2006te}, both at weak coupling by means of perturbation theory, and at strong coupling by means of holography (see also \cite{Teaney:2006nc} for computations of the relevant spectral functions at strong coupling from holography). Photon production rates in $\mathcal{N} = 4$ SYM including leading order corrections in the 't Hooft coupling $\lambda_t$ were considered in Ref.'s \cite{Hassanain:2011ce,Hassanain:2012uj} (in the last one, leading order corrections in the number of colors $N_c$ were also calculated specifically for the electric conductivity). Extensions of these results to anisotropic deformations of $\mathcal{N} = 4$ SYM were also considered in the literature, including cases where the anisotropy is due to the presence of an external magnetic field \cite{Mamo:2013efa}, scenarios where the anisotropy is driven by a non-trivial bulk axion field \cite{Patino:2012py,Jahnke:2013rca,Misobuchi:2015ioa}, and also holographic setups comprising both effects \cite{Wu:2013qja,Muller:2013ila}.

The thermal photon production rate for a non-conformal QGP at zero baryon chemical potential in a similar holographic setup to the one we are going to consider here was recently obtained in \cite{Yang:2015bva}. We shall discuss the qualitative similarities and the quantitative differences we found between our results and the results obtained in \cite{Yang:2015bva} for the thermal photon spectrum at $\mu_B=0$.

This work is organized as follows: in Section \ref{sec:model} we briefly review the main features of the phenomenological EMD model \cite{Rougemont:2015wca} which will be used throughout this work. In Section \ref{sec:transport} we setup the flow equations used for computing the retarded propagator of the electric current (needed in the calculation of the thermal spectra) and apply these flow equations to obtain different electric charge transport coefficients, namely the DC and AC electric conductivities and the electric charge diffusion constant as functions of $T$ and $\mu_B$. We also calculate the electric charge susceptibility as a function of $T$ and $\mu_B$. In Section \ref{sec:spectra} we compute the thermal photon and dilepton spectra at zero and nonzero baryon chemical potential. We finish in Section \ref{sec:conclusion} by presenting a discussion of the main results of the present work, and in particular how our holographic calculations constraint the thermal photon and dilepton production rates in a strongly coupled QGP to be at least one order of magnitude below perturbative estimates.

We make use of the following conventions throughout this paper: we work with natural units where $\hbar = k_B = c = 1$ and the metric signature used is $(-,+,+,+,+)$. Uppercase Latin indices $M,N,\ldots$ run over the whole bulk of the holographic spacetime, $M,N,\ldots = r,t,x,y,z$; lowercase Greek indices $\mu,\nu,\ldots$ run over the boundary field theory coordinates, $\mu,\nu,\ldots = t,x,y,z$; lowercase latin indices $i,j,\ldots$ run over the spatial coordinates of the field theory coordinates, $i,j,\ldots = x,y,z$.

\section{The holographic model}
\label{sec:model}

In this Section we are going to sketch the main features of the holographic model \cite{Rougemont:2015wca} used in the present work. For a detailed account on the construction of this model, its thermodynamic properties and many transport observables associated to the energy loss of light and heavy quarks traversing a hot and baryon rich medium, we refer the reader to consult \cite{Rougemont:2015wca}; see also \cite{Rougemont:2015ona} for results concerning many transport coefficients associated to the diffusion of baryon charge. Here, we shall only review the essential results in order to keep this work self-contained and also to introduce the scalings needed to compute the physical observables in the so-called numerical coordinates, as we shall discuss in Section \ref{sec:scaling}.

For results concerning a large set of first and second order hydrodynamic transport coefficients at $\mu_B = 0$ we refer the reader to Ref. \cite{Finazzo:2014cna}. For a calculation of the crossover temperature dependence on an external magnetic field at $\mu_B = 0$, see \cite{Rougemont:2015oea}.

\subsection{Einstein-Maxwell-Dilaton action}

The bottom-up holographic model considered here is described by a five dimensional EMD action,
\begin{align}
S&=\frac{1}{16\pi G_5} \int_{\mathcal{M}_5}d^5x \, \sqrt{-g} \left[\mathcal{R}-\frac{1}{2}(\partial_M \phi)^2-V(\phi) -\frac{f_{B}(\phi)}{4}F_{MN}^2\right]+S_{\textrm{GHY}}+S_{\textrm{CT}},
\label{eq:EMDaction}
\end{align}
where $G_5$ is the five dimensional Newton's constant, $g_{MN}$ is the bulk metric, $\mathcal{R}$ is the corresponding Ricci scalar, $F_{MN}$ is the field strength tensor for the Maxwell field $A_{M}$ whose boundary value of its temporal component gives the baryon chemical potential $\mu_B$ at the boundary gauge field theory, $\phi$ is the dilaton, a real scalar field with a non-trivial profile in the bulk which, under an adequate choice for its potential $V(\phi)$, triggers a holographic renormalization group flow breaking the conformal symmetry in the infrared (IR) regime of the boundary gauge theory, and $f_{B} (\phi)$ is the Maxwell-Dilaton baryon coupling function. The Newton's constant $G_5$ and the dilaton potential $V(\phi)$ shall be dynamically fixed by demanding that the holographic equation of state at $\mu_B=0$ matches the corresponding lattice results obtained in Ref. \cite{Borsanyi:2012cr} for $(2+1)$-flavor QCD with physical quark masses, while the Maxwell-Dilaton baryon coupling $f_B(\phi)$ will be dynamically fixed by matching the lattice results for the baryon susceptibility at $\mu_B=0$ \cite{Borsanyi:2011sw}.

Note the action \eqref{eq:EMDaction} also comprises two boundary terms, where $S_{\textrm{GHY}}$ is the Gibbons-Hawking-York action \cite{York:1972sj,Gibbons:1976ue} needed to establish a well-posed variational problem with Dirichlet boundary condition for the metric field, and $S_{\textrm{CT}}$ is the counterterm action constructed by means of the holographic renormalization procedure \cite{Henningson:1998gx,deHaro:2000xn,Skenderis:2002wp,Papadimitriou:2011qb,Lindgren:2015lia}, which is required to cancel the ultraviolet (UV) divergences of the on-shell action. These two boundary actions only contribute to the real part of Green's functions, but since in the calculations to be carried out here we shall be only concerned with the imaginary part of Green's functions, which are free of divergences, these two boundary actions will not be needed, therefore, we drop them out.

As detailed discussed in Ref.'s \cite{Rougemont:2015wca,DeWolfe:2010he,DeWolfe:2011ts,Gubser:2008ny,Gubser:2008yx,Rougemont:2015ona}, the EMD model considered here lies within a class of bottom-up models regarded as effective holographic duals \emph{mimicking} in a quantitative way at least part of the $(2+1)$-flavor QCD properties as manifest in the behavior of some physical observables. In fact, even though our simple EMD construction does not explicitly introduce fundamental flavors at the dual boundary quantum field theory, adequate choices for $V(\phi)$ and $f_{B} (\phi)$ (to be reviewed in Section \ref{sec:thermo}) are able to \emph{emulate} in a quantitative way many of the properties of the $(2+1)$-flavor QGP around the crossover transition as simulated on the lattice. Indeed, in Ref. \cite{Rougemont:2015wca}, a highly non-trivial achievement was the quantitative agreement between the holographic equation of state and the corresponding lattice data \cite{Borsanyi:2012cr} around the crossover transition for baryon chemical potentials up to 400 MeV. Moreover, in Ref. \cite{Rougemont:2015ona}, very remarkably, this same model was also shown to quantitatively describe the splitting between the second and fourth order baryon susceptibilities in the deconfined phase, in agreement with a very recent lattice calculation \cite{Bellwied:2015lba}. Since first principle QCD results for real time observables at finite baryon chemical potential near the crossover transition may be out of the reach of lattice calculations for quite a long time, these results constitute a very good motivation to investigate non-equilibrium phenomena using the present bottom-up holographic model.

\subsection{Ansatz in standard coordinates}
\label{sec:standard}

We take the following ansatze for the EMD fields
\begin{align}
d\tilde{s}^2=e^{2\tilde{A}(\tilde{r})}\left[-\tilde{h}(\tilde{r})d\tilde{t}^2+d\vec{\tilde{x}}^2\right]+ \frac{d\tilde{r}^2}{\tilde{h}(\tilde{r})},\,\,\,\tilde{\phi}=\tilde{\phi}(\tilde{r}),\,\,\, \tilde{A}=\tilde{A}_M d\tilde{x}^M=\tilde{\Phi}(\tilde{r})d\tilde{t},
\label{eq:metricansatz}
\end{align}
where $\tilde{r}$ is the radial holographic coordinate and $\tilde{A}(\tilde{r})$, $\tilde{h}(\tilde{r})$, $\phi(\tilde{r})$, and $\tilde{\Phi} (\tilde{r})$ are functions of $\tilde{r}$ only. The function $\tilde{h}(\tilde{r})$ is the blackening factor: in order to have a black brane in the IR portion of the bulk geometry, we impose that the blackening function has a simple zero at $\tilde{r}=\tilde{r_H}$, which is the radial location of the spatially extended black brane horizon. This ansatz preserves the $SO(3)$ rotational symmetry but breaks Lorentz invariance whenever $\tilde{h}(\tilde{r}) \neq 1$, being consistent with a finite temperature formalism. We have consistently set to zero every component of $\tilde{A}_M$, except for $\tilde{A}_t$, which is taken to be dual to the baryon chemical potential $\mu_B$ at the boundary gauge theory. In this coordinate system, the boundary is at $\tilde{r} \to \infty$ and, near the boundary, $\tilde{A}$, $\tilde{h}$, $\tilde{\phi}$, and $\tilde{\Phi}$ have the following UV asymptotic behavior \cite{DeWolfe:2010he,DeWolfe:2011ts}
\begin{align}
\tilde{A}(\tilde{r})&=\tilde{r}+\mathcal{O}\left(e^{-2\nu\tilde{r}}\right),\nonumber\\
\tilde{h}(\tilde{r})&=1+\mathcal{O}\left(e^{-4\tilde{r}}\right),\nonumber\\
\tilde{\phi}(\tilde{r})&=e^{-\nu\tilde{r}}+\mathcal{O}\left(e^{-2\nu\tilde{r}}\right),\nonumber\\
\tilde{\Phi}(\tilde{r})&=\tilde{\Phi}_0^{\textrm{far}}+\tilde{\Phi}_2^{\textrm{far}}e^{-2\tilde{r}}+ \mathcal{O}\left(e^{-(2+\nu)\tilde{r}}\right),
\label{eq:asymptoticansatz}
\end{align}
where $\nu \equiv d - \Delta$, $d = 4$ is the number of dimensions of the boundary gauge theory and $\Delta$ is the scaling dimension of a relevant operator dual to $\tilde{\phi}$. Also, $\tilde{\Phi}_0^{\textrm{far}}$ and $\tilde{\Phi}_2^{\textrm{far}}$ are constants. We remark that the metric is asymptotically $\mathrm{AdS}_5$ and the gauge theory has, therefore, a strongly coupled UV fixed point.

In Eq. \eqref{eq:metricansatz}, we followed the same notation of Ref.'s \cite{Rougemont:2015wca,DeWolfe:2010he,DeWolfe:2011ts,Rougemont:2015ona}, where the tildes are employed to express the EMD fields in the so-called \emph{standard coordinates}, where the blackening function goes to unity at the boundary. In these coordinates, one may use standard holographic expressions to obtain the physical observables of the theory. For instance, the entropy density follows from the Bekenstein-Hawking's formula \cite{Bekenstein:1973ur,Hawking:1974sw}
\begin{align}
\hat{s}=\frac{2\pi}{\kappa^2}e^{3\tilde{A}(\tilde{r}_H)},
\label{eq:entropy}
\end{align}
where $\kappa^2 \equiv 8 \pi G_5$, while the temperature $\hat{T}$ is given by the Hawking's temperature of the black brane horizon
\begin{align}
\hat{T}=\frac{\sqrt{-g'_{\tilde{t}\tilde{t}} g^{\tilde{r}\tilde{r}}\,'}}{4\pi}\biggr|_{\tilde{r}=\tilde{r}_H}= \frac{e^{\tilde{A}(\tilde{r}_H)}}{4\pi}|\tilde{h}'(\tilde{r}_H)|.
\label{eq:temperature}
\end{align}
Also, according to the holographic dictionary, the baryon chemical potential $\hat{\mu}_B$ and the baryon charge density $\hat{\rho}_B$ are given by
\begin{align}
\hat{\mu}_B=\lim_{\tilde{r}\rightarrow\infty}\tilde{A}_t(\tilde{r})=\tilde{\Phi}_0^{\textrm{far}}\,\,\,\,\,\textrm{and}\,\,\,\,\,
\hat{\rho}_B=\lim_{\tilde{r}\rightarrow\infty} \frac{\partial\mathcal{L}}{\partial\left(\partial_{\tilde{r}}\tilde{A}_t\right)}
=-\frac{\tilde{\Phi}_2^{\textrm{far}}}{\kappa^2}.
\label{eq:chempot}
\end{align}
In Eq.'s \eqref{eq:entropy} to \eqref{eq:chempot}, the hat on the thermodynamic observables is used to indicate that they are being measured in powers of the inverse of the AdS radius $L$, which we have set to unity. In Sections \ref{sec:scaling} and \ref{sec:thermo}, we are going to discuss how one may express these observables in natural units, which shall then be denoted without the hat.

\subsection{Ansatz in numerical coordinates}
\label{sec:numerical}

As detailed in Ref. \cite{Rougemont:2015wca}, in order to numerically solve the EMD equations of motion, one needs first to take a near-horizon Taylor expansion for the unknown functions $\tilde{A}(\tilde{r})$, $\tilde{h}(\tilde{r})$, $\phi(\tilde{r})$, and $\tilde{\Phi} (\tilde{r})$, determine the on-shell Taylor coefficients of these expansions and then integrate the equations of motion from the horizon up to the boundary. Taking a second order Taylor expansion, there are 12 Taylor coefficients to be determined in order to initialize the numerical integrations. Two of these coefficients are the initial conditions corresponding to the horizon values of the dilaton field and the first derivative of the Maxwell field. Numerical geometries obtained with different choices for these two initial conditions translate into different thermodynamic states $(T,\mu_B,s,\rho_B)$ at the boundary gauge field theory. In order to perform the numerical integrations, we must also specify numerical values for the 10 remaining Taylor coefficients in the second order near-horizon expansions. One of them, corresponding to the horizon value of the blackening function, vanishes by definition at the horizon. Furthermore, since $dt$ has infinite norm at the horizon, the horizon value of the Maxwell field must be set to zero by consistency. We may completely fix the gauge freedom present in the ansatz \eqref{eq:metricansatz} by conveniently rescaling the bulk coordinates such as to place the horizon location at $r_H=0$ and also taking $h'(r_H) = 1$ and $A(r_H) = 0$, where the new coordinates obtained after these rescalings are the so-called \emph{numerical coordinates}, which we denote without the tildes. The 6 remaining Taylor coefficients in the second order near-horizon expansions are then determined on-shell as functions of the initial conditions $\phi_0 \equiv \phi(r_H)$ and $\Phi_1 \equiv \Phi'(r_H)$.

In the numerical coordinates, the form of the ansatz for the EMD fields reads as follows
\begin{align}
ds^2=e^{2A(r)}\left[-h(r)dt^2+d\vec{x}^2\right]+\frac{dr^2}{h(r)},\,\,\,\phi=\phi(r),\,\,\,A=A_M dx^M=\Phi(r)dt,
\label{eq:metricansatz2}
\end{align}
with the bulk fields satisfying the following UV asymptotic behavior near the boundary \cite{DeWolfe:2010he,DeWolfe:2011ts}
\begin{align}
A(r)&=\alpha(r)+\mathcal{O}\left(e^{-2\nu\alpha(r)}\right);\,\,\,\alpha(r)= A_{-1}^{\textrm{far}}r+A_0^{\textrm{far}},\nonumber\\
h(r)&=h_0^{\textrm{far}}+\mathcal{O}\left(e^{-4\alpha(r)}\right),\nonumber\\
\phi(r)&=\phi_A e^{-\nu\alpha(r)}+\mathcal{O}\left(e^{-2\nu\alpha(r)}\right),\nonumber\\
\Phi(r)&=\Phi_0^{\textrm{far}}+\Phi_2^{\textrm{far}}e^{-2\alpha(r)}+ \mathcal{O}\left(e^{-(2+\nu)\alpha(r)}\right).
\label{eq:asympnum}
\end{align}

\subsection{Scaling relations}
\label{sec:scaling}

The equations of motion following from the EMD action \eqref{eq:EMDaction} are solved in the numerical coordinates with the ansatz \eqref{eq:metricansatz2}, but the physical observables are computed in the standard coordinates discussed in Section \ref{sec:standard}. Thus, one needs to write down the dictionary relating these two coordinate systems. This is accomplished by comparing the UV asymptotic behaviors \eqref{eq:asymptoticansatz} and \eqref{eq:asympnum} with the requirement that $d\tilde{s}^2 = ds^2$, $\tilde{\phi}(\tilde{r}) = \phi(r)$, and $\tilde{\Phi}(\tilde{r})d\tilde{t}=\Phi(r)dt$, from which one obtains the following scaling relations
\begin{align}
\tilde{r}&=\frac{r}{\sqrt{h_0^{\textrm{far}}}}+A_0^{\textrm{far}}-\ln(\phi_A^{1/\nu}),\nonumber\\
\tilde{A}(\tilde{r})&=A(r)-\ln(\phi_A^{1/\nu}),\nonumber\\
\vec{\tilde{x}}&=\phi_A^{1/\nu}\vec{x},\nonumber\\
\tilde{t}&=\phi_A^{1/\nu}\sqrt{h_0^{\textrm{far}}}\,t,\nonumber\\
\tilde{h}(\tilde{r})&=\frac{h(r)}{h_0^{\textrm{far}}},\nonumber\\
\tilde{\Phi}(\tilde{r})&=\frac{\Phi(r)}{\phi_A^{1/\nu}\sqrt{h_0^{\textrm{far}}}},\nonumber\\
\tilde{\Phi}_0^{\textrm{far}}&=\frac{\Phi_0^{\textrm{far}}}{\phi_A^{1/\nu}\sqrt{h_0^{\textrm{far}}}},\nonumber\\
\tilde{\Phi}_2^{\textrm{far}}&=\frac{\Phi_2^{\textrm{far}}}{\phi_A^{3/\nu}\sqrt{h_0^{\textrm{far}}}}.
\label{eq:scalingrelations}
\end{align}

By applying the scaling relations \eqref{eq:scalingrelations} into Eq.'s \eqref{eq:entropy} to \eqref{eq:chempot}, one may write down expressions for the thermodynamic variables $\hat{T}$, $\hat{\mu}_B$, $\hat{s}$, and $\hat{\rho}_B$ in the numerical coordinates
\begin{eqnarray}
\hat{T}&=&\frac{1}{4\pi\phi_A^{1/\nu}\sqrt{h_0^{\textrm{far}}}},\nonumber\\
\hat{\mu}_B&=&\frac{\Phi_0^{\textrm{far}}}{\phi_A^{1/\nu}\sqrt{h_0^{\textrm{far}}}},\nonumber\\
\hat{s}&=&\frac{2\pi}{\kappa^2\phi_A^{3/\nu}},\nonumber\\
\hat{\rho}_B&=&-\frac{\Phi_2^{\textrm{far}}}{\kappa^2\phi_A^{3/\nu}\sqrt{h_0^{\textrm{far}}}}.
\label{eq:physicalthings}
\end{eqnarray}

As discussed in Ref.'s \cite{Rougemont:2015wca,Rougemont:2015oea}, one may convert to natural units the expression for any physical observable $\hat{X}$ with mass dimension $p$ measured in units of the $p$-th inverse power of the AdS radius by employing a scaling factor $\Lambda$ with mass dimension 1 measured in MeV (for instance). We denote by $X = \Lambda^p \hat{X}$ the expression for the physical observable $X$ measured in units of MeV$^p$. In the next Section, we shall present a procedure to fix the dimensionful scaling factor $\Lambda$. We remark that by using a single scaling factor $\Lambda$ we do not augment the number of free parameters of the holographic model, since we are basically exchanging the freedom of choosing the AdS radius by the freedom of choosing $\Lambda$, and also, this prescription respects the fact that any dimensionless ratio must correspond to a pure number independent of the units employed.

In short, the procedure used to compute a physical observable $X$ in the present holographic model comprises the following general steps: (a) one first derives a holographic expression for $\hat{X}$ in the standard coordinates, (b) then one applies the scaling relations \eqref{eq:scalingrelations} to rewrite $\hat{X}$ in the numerical coordinates, which are the coordinates in terms of which the numerical solutions of the EMD equations of motion are obtained, (c) and, finally, one obtains $X$ from $\hat{X}$ by using the scaling factor $\Lambda$ as discussed above.

\subsection{Model parameters and thermodynamics}
\label{sec:thermo}

\begin{figure}[htp!]
\begin{centering}
\includegraphics[width=0.46\textwidth]{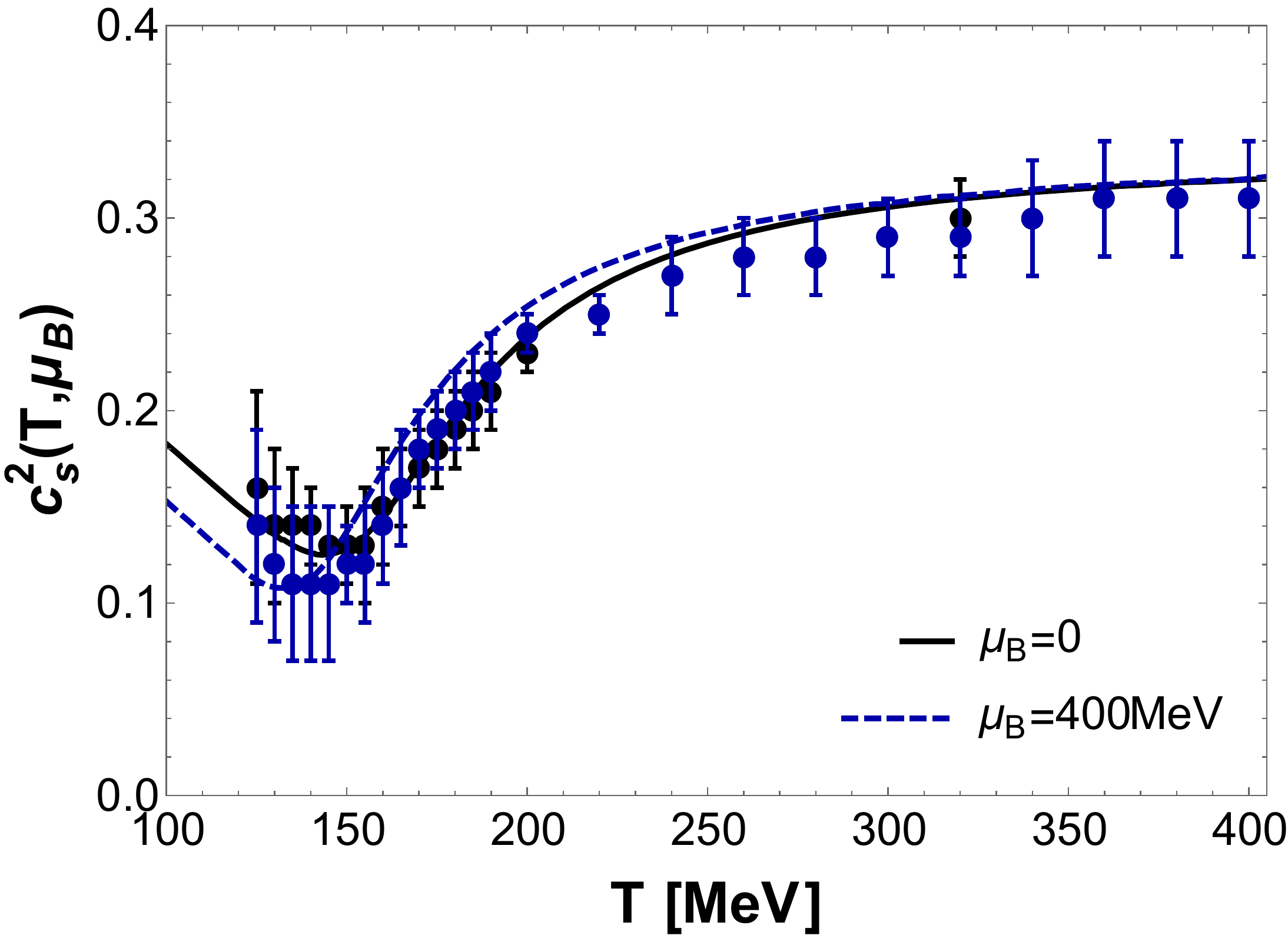}
\hskip0.03\textwidth
\includegraphics[width=0.46\textwidth]{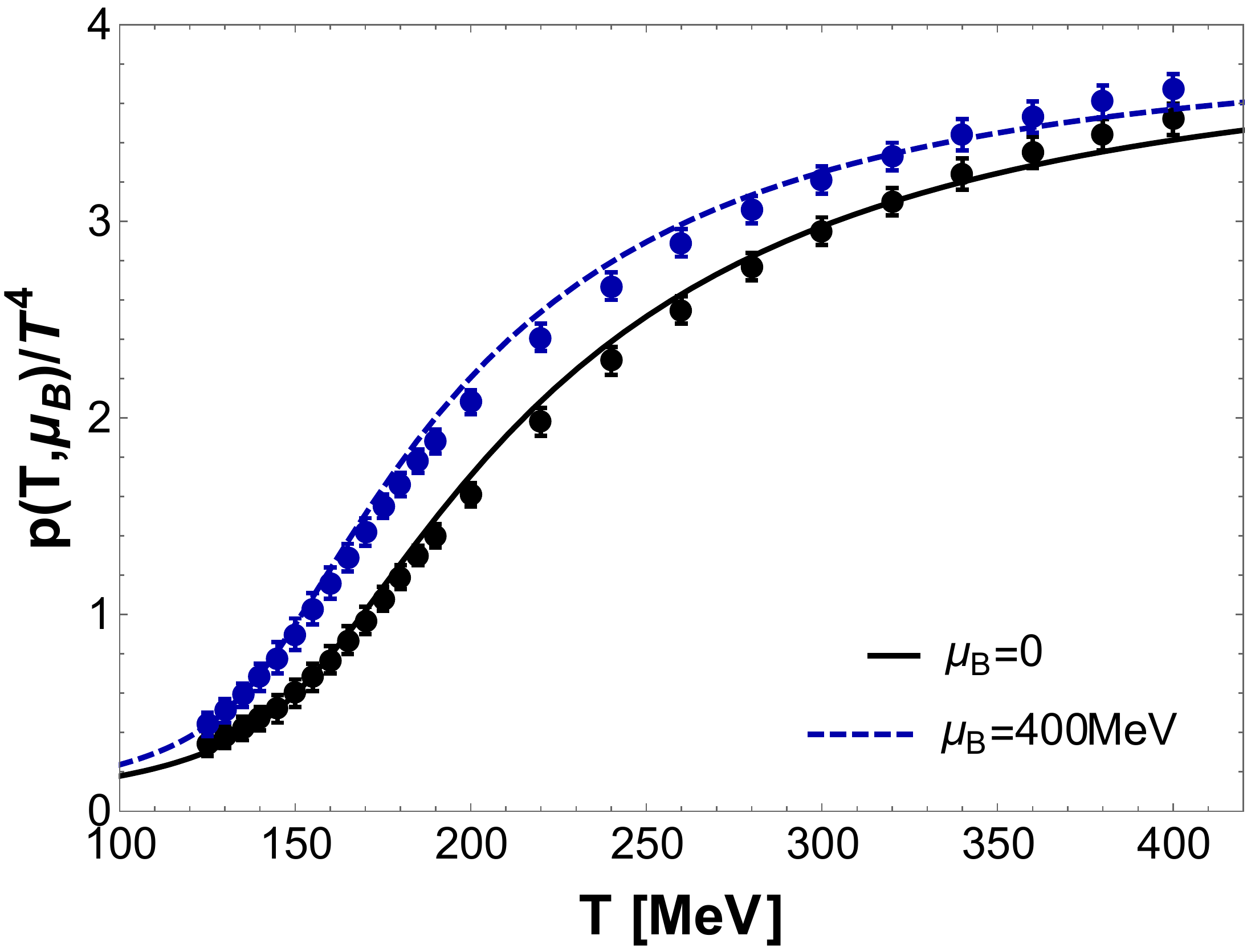}
\par\end{centering}
\caption{(Color online) Speed of sound squared $c_s^2$ and normalized pressure $p/T^4$ as functions of the temperature $T$ at different fixed values of the baryon chemical potential $\mu_B$. The points with error bars correspond to lattice data for $(2+1)$-flavor QCD with physical quark masses from Ref. \cite{Borsanyi:2012cr}.\label{fig:thermodynamics}}
\end{figure}

\begin{figure}[htp!]
\begin{centering}
\includegraphics[scale=0.52]{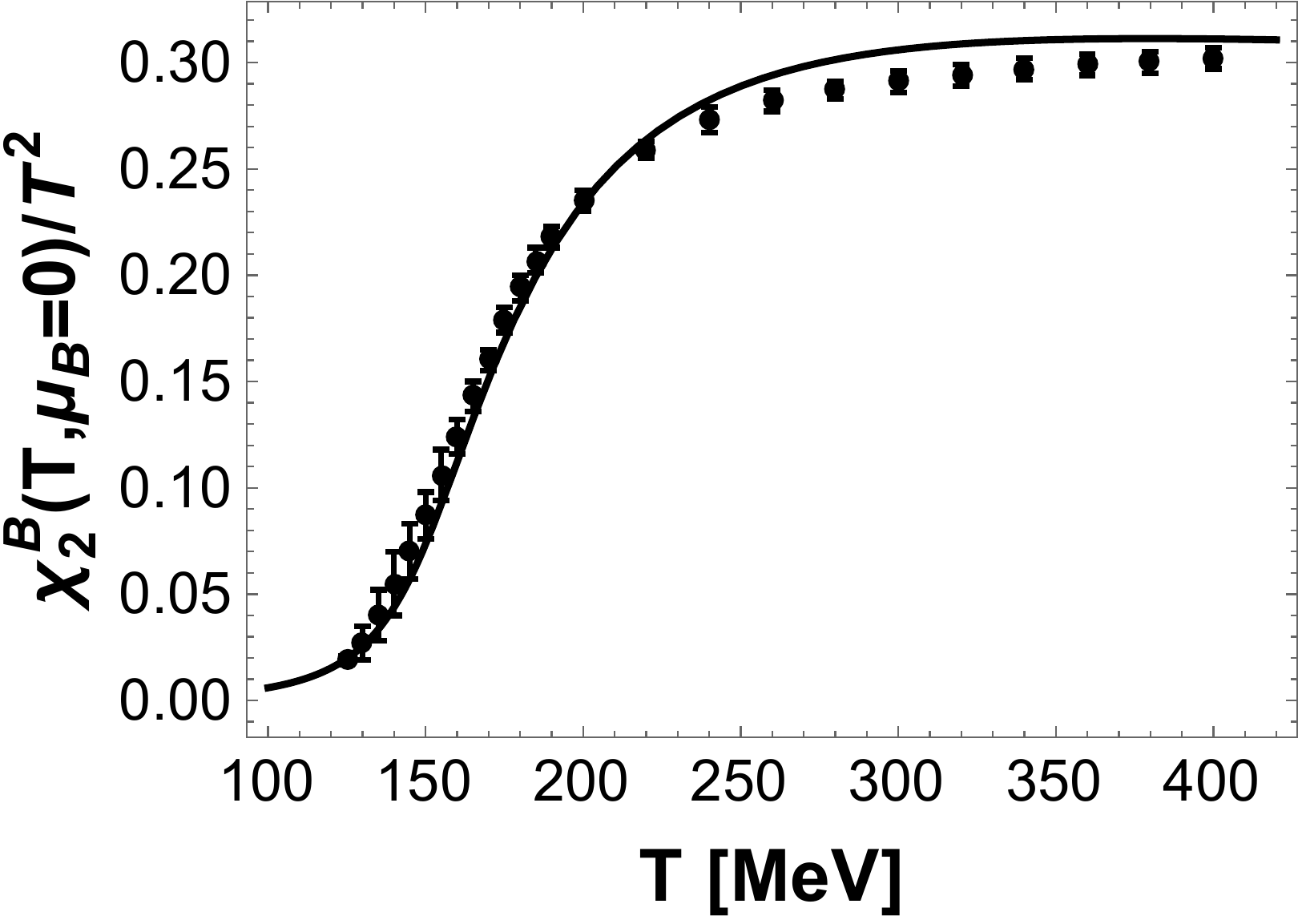}
\par\end{centering}
\caption{Normalized baryon susceptibility at zero baryon chemical potential as a function of the temperature. The points with error bars correspond to lattice data for $(2+1)$-flavor QCD with physical quark masses from Ref. \cite{Borsanyi:2011sw}.}
\label{fig:baryonchi2}
\end{figure}

As detailed discussed in Ref. \cite{Rougemont:2015wca}, the free parameters of the EMD action \eqref{eq:EMDaction} may be dynamically fixed be solving the EMD equations of motion with the requirement that the holographic results for the equation of state and the baryon susceptibility at $\mu_B=0$ fit the corresponding lattice results. More specifically, one may fix the dilaton potential $V(\phi)$ and the gravitational constant $G_5$ by fitting lattice data \cite{Borsanyi:2012cr} for the speed of sound squared $c_s^2(T,\mu_B=0)$ and the normalized pressure $p(T,\mu_B=0)/T^4$, respectively\footnote{This prescription was firstly put forward in Ref. \cite{Ficnar:2010rn}.}, while the Maxwell-Dilaton baryon coupling $f_B(\phi)$ is fixed by fitting lattice data \cite{Borsanyi:2011sw} for the normalized baryon susceptibility $\chi_2^B(T,\mu_B=0)/T^2=T^{-2}[\partial\rho_B/\partial\mu_B]_{\mu_B=0}$. We found that the following set of parameters
\begin{align}
V(\phi)&=-12\cosh(0.606\,\phi)+0.703\,\phi^2-0.1\,\phi^4+0.0034\,\phi^6;\,\,\,\kappa^2 = 8\pi G_5 = 12.5,\label{eq:potential}\\
f(\phi)&=\frac{\textrm{sech}(1.2\,\phi-0.69)}{3\,\textrm{sech}(0.69)}+\frac{2}{3}\,e^{-100\,\phi},\label{eq:fBpar}
\end{align}
provide a good fit to lattice data at $\mu_B=0$, as shown in Fig.'s \ref{fig:thermodynamics} and \ref{fig:baryonchi2}. In Fig. \ref{fig:thermodynamics}, one also observes that the holographic predictions for $c_s^2(T,\mu_B)$ and $p(T,\mu_B)/T^4$ at the highest value of the baryon chemical potential reached in the current BES at RHIC, namely $\mu_B=400$ MeV, are in good quantitative agreement with the corresponding lattice data \cite{Borsanyi:2012cr}.\footnote{The scaling dimension of the relevant operator dual to the dilaton field obtained from the potential in Eq. \eqref{eq:potential} is $\Delta \approx 3$ \cite{Finazzo:2014cna,Rougemont:2015wca}, corresponding to a squared dilaton mass of $m^2 \approx -3$, which satisfies the Breitenlohner-Freedman bound \cite{Breitenlohner:1982jf,Breitenlohner:1982bm} for a scalar field on asymptotically AdS$_5$ spaces.}

We note that the dimensionful scaling factor $\Lambda\approx 831$ MeV used in these holographic calculations was obtained by matching the black brane Hawking's temperature in the dip of the holographic speed of sound squared at $\mu_B=0$ with the corresponding lattice result. We also remark that the same parametrization for the dilaton potential used in Eq. \eqref{eq:potential} was already successfully employed in Ref.'s \cite{Rougemont:2015wca,Rougemont:2015ona,Finazzo:2014cna,Rougemont:2015oea,Janik:2015waa} to compute several physical observables of interest to the QGP phenomenology.

\section{Electric charge transport phenomena at $\mu_B=0$ and $\mu_B\neq 0$}
\label{sec:transport}

\subsection{Bulk electric charge sector}
\label{sec:bulkcharge}

In this Section, we shall analyze electric charge transport phenomena in the holographic model defined in the last Section, at both zero and nonzero baryon chemical potential, extending the previous work \cite{Finazzo:2013efa} to the $(T,\mu_B)$-plane.

In QCD one usually makes the assumption that charm quarks are of no relevance in the crossover transition, in which case there are three independent chemical potentials associated with the global charge conservation of $(2+1)$ flavors, $\mu_u$, $\mu_d$, and $\mu_s$. These three quark chemical potentials are usually reorganized into the baryon chemical potential $\mu_B$, the strangeness chemical potential $\mu_S$, and the electric charge chemical potential $\mu_Q$. By means of the holographic dictionary, one is instructed to associate to each of these $U(1)$ globally conserved charges a different Maxwell field in the bulk. Moreover, by taking into account the hierarchy among these different chemical potentials, as observed in the experimental conditions produced in current heavy ion collisions at RHIC, one concludes that, as a first approximation, it is reasonable to take only a non-vanishing $\mu_B$, setting $\mu_S=\mu_Q=0$. In fact, from the results obtained in Ref.'s \cite{Bazavov:2012vg,Borsanyi:2013hza,Karsch:2010ck}, one sees that, roughly speaking, $\mu_B \sim 10 \mu_S \sim 100 \mu_Q$ at RHIC. With such an observation in sight, we are going to study electric charge transport in the present holographic model by taking a probe approximation where the electric charge sector is affected by the baryon charge sector, but the former does not backreact on the latter. In this way, one may consider fluctuations $B_M$ (with associated field strength tensor $G_{MN}$) of a second bulk Maxwell field $\mathcal{B}_M$ (with vanishing background value, corresponding to set $\mu_Q=0$) sourcing the electric current operator at the boundary gauge theory and write down the following probe action on top of the numerical backgrounds constructed over the $(T,\mu_B)$-plane as discussed in the last Section
\begin{equation}
\label{eq:fluctuation}
S_Q = - \frac{1}{16\pi G_5} \int_{\mathcal{M}_5}d^5x \, \sqrt{-g}\frac{f_{Q}(\phi)}{4}G_{MN}^2,
\end{equation}
where $f_Q(\phi)$ is the Maxwell-Dilaton electric coupling (to be fixed in Section \ref{sec:electricsusceptibility}). The bulk perturbation $B_{M}$ in the probe action \eqref{eq:fluctuation} is dual to boundary gauge theory fluctuations of the conserved electric charge in the $(T,\mu_B)$-plane at $\mu_Q=0$.

\subsection{Flow equations}
\label{sec:flow}

In order to compute electric charge transport coefficients and the thermal photon and dilepton spectra from holography using the bulk fluctuations of the electric charge sector described by the action \eqref{eq:fluctuation}, we employ the flow equations derived in Ref. \cite{Iqbal:2008by}, which provide an useful implementation of the real time gauge/gravity prescription. The aim of this method is to compute the imaginary part of the thermal retarded propagator of the boundary gauge theory current operator defined by
\begin{equation}
G^{\mu \nu}_R(\tilde{k}^\mu) =-i\int_{\mathbb{R}^{1,3}} d^4 \tilde{x}\,e^{-i \tilde{k}_\mu\tilde{x}^\mu}\theta(\tilde{t})\left\langle \left[J_Q^{\mu}(\tilde{k}^\mu),J_Q^{\nu}(0)\right]\right\rangle_{T},
\end{equation}
in terms of which one defines the spectral function as follows
\begin{equation}
\chi^{\mu \nu}(\tilde{k}^\mu) = - 2 \mathrm{Im}[G^{\mu \nu}_R(\tilde{k}^\mu)],
\label{eq:spectral}
\end{equation}
where $\tilde{k}^{\mu} = (\tilde{\omega},\vec{\tilde{k}})$ is the four-momentum of the photon at the boundary gauge theory. From the above definition for the spectral function and the Kubo's relation between the conductivity tensor and the retarded current propagator obtained in linear response theory \cite{Iqbal:2008by},
\begin{equation}
\sigma^{\mu \nu}(\tilde{k}^\mu) = -\frac{G^{\mu \nu}_R(\tilde{k}^\mu)}{i\tilde{\omega}},
\label{eq:Kubo}
\end{equation}
it follows that the trace of the spectral function may be written as below
\begin{equation}
\chi^\mu_\mu(\tilde{k}^\mu) = 2\tilde{\omega}\mathrm{Re}[\sigma^\mu_\mu(\tilde{k}^\mu)].
\label{eq:trace}
\end{equation}

By taking the spatial momentum of the photon along the $z$-direction, we define the following quantities, whose boundary values give the longitudinal and transverse conductivities, respectively \cite{Iqbal:2008by}
\begin{align}
\sigma_L(\tilde{r})\equiv\sigma^{zz}(\tilde{r})=\frac{\tilde{j}^z}{i\tilde{\omega}\tilde{B}_z}\,\,\,\,\textrm{and}\,\,\,\,
\sigma_T(\tilde{r})\equiv\sigma^{xx}(\tilde{r})=\sigma^{yy}(\tilde{r})=\frac{\tilde{j}^x}{i\tilde{\omega}\tilde{B}_x},
\label{eq:defsigmas}
\end{align}
where $\tilde{j}^M = \delta S_Q/\delta(\partial_{\tilde{r}}\tilde{B}_M)$ is the radial momentum conjugate to the perturbation $\tilde{B}_M$. By taking $\tilde{k}^{\mu} = (\tilde{\omega},0,0,\tilde{k})$, the flow equations for $\sigma_L$ and $\sigma_T$ read \cite{Iqbal:2008by}
\begin{align}
\partial_{\tilde{r}}\sigma_L & = \frac{i \tilde{\omega} e^{-\tilde{A}(\tilde{r})}}{\tilde{h}(\tilde{r})} \left[ \frac{\sigma_L^2(\tilde{r})}{\tilde{\Sigma}(\tilde{r})} \left(1 - \tilde{h}(\tilde{r}) \frac{\tilde{k}^2}{\tilde{\omega}^2} \right) - \tilde{\Sigma}(\tilde{r}) \right], \label{eq:flow1} \\
\partial_{\tilde{r}}\sigma_T & = \frac{i \tilde{\omega} e^{-\tilde{A}(\tilde{r})}}{\tilde{h}(\tilde{r})} \left[ \frac{\sigma_T^2(\tilde{r})}{\tilde{\Sigma}(\tilde{r})} - \tilde{\Sigma}(\tilde{r})\left(1 - \tilde{h}(\tilde{r}) \frac{\tilde{k}^2}{\tilde{\omega}^2} \right) \right], \label{eq:flow2}
\end{align}
where we defined
\begin{equation}
\tilde{\Sigma}(\tilde{r}) = \frac{f_Q(\tilde{\phi}(\tilde{r})) e^{\tilde{A}(\tilde{r})}}{16\pi G_5}.
\label{eq:flow5}
\end{equation}
Since the flow equations \eqref{eq:flow1} and \eqref{eq:flow2} are first order differential equations, in order to solve them we need to specify initial data for $\sigma_L$ and $\sigma_T$ at the horizon. Since the blackening function vanishes at the horizon, we see from the structure of the flow equations that, by requiring regularity of the solutions at the horizon, the initial data required to start the numerical integration of the flow equations are given by \cite{Iqbal:2008by}
\begin{equation}
\sigma_L(\tilde{r}_H) = \sigma_T(\tilde{r}_H) = \tilde{\Sigma}(\tilde{r}_H).
\end{equation}

In order to rewrite the above equations in the numerical coordinates, one needs to know also how to translate the four-momentum $\tilde{k}^\mu=(\tilde{\omega},0,0,\tilde{k})$ from standard to numerical coordinates. This can be done by noting that $\tilde{k}_\mu \tilde{x}^{\mu}$ must have the same value in both coordinates, since it is a Lorentz scalar. Then, by using the scaling relations \eqref{eq:scalingrelations} and also the expression for the temperature in the numerical coordinates \eqref{eq:physicalthings}, one obtains
\begin{align}
\tilde{\omega}=\frac{\omega}{\phi_A^{1/{\nu}} \sqrt{h_0^{\textrm{far}}}}=4\pi T\omega\,\,\,\,\textrm{and}\,\,\,\,
\tilde{k}=\frac{k}{\phi_A^{1/{\nu}}}=4\pi Tk\sqrt{h_0^{\textrm{far}}}.
\label{eq:scalingmomentum}
\end{align}
The flow equations written in the numerical coordinates are simply given by Eq.'s \eqref{eq:flow1} and \eqref{eq:flow2} without the tildes. On the other hand, Eq. \eqref{eq:flow5} reads as follows in the numerical coordinates
\begin{equation}
\label{eq:sigma}
\Sigma(r) = \frac{f_Q(\phi)}{16 \pi G_5} \frac{e^{A(r)}}{\phi_A^{1/{\nu}}}.
\end{equation}

\subsection{Electric susceptibility and electric charge transport}
\label{sec:transportcoefficients}

\subsubsection{Electric charge susceptibility}
\label{sec:electricsusceptibility}

In order to dynamically fix the Maxwell-Dilaton electric coupling $f_Q(\phi)$, we need first to obtain a holographic formula for the electric charge susceptibility $\chi^Q_2 = \partial \rho_Q/\partial \mu_Q$ at $\mu_B=\mu_Q=0$, where $\rho_Q$ is the electric charge density. The derivation of this formula can be done directly from the flow equations by considering the diffusion process in the longitudinal channel $\sigma_L$, as discussed in Ref. \cite{Iqbal:2008by}. Alternatively, one may also consider the steps followed in Section 5.1 of Ref. \cite{DeWolfe:2010he} (see also Section 3.1 of Ref. \cite{Rougemont:2015wca}). The result in the numerical coordinates reads as follows
\begin{equation}
\label{eq:susceptibility2}
\frac{\chi^Q_2}{T^2}(\mu_B=\mu_Q = 0) = \frac{1}{16\pi^2} \frac{s}{T^3} \frac{1}{f_Q(0)\int_{r_H}^\infty dr\, e^{-2A(r)}f_Q^{-1}(\phi(r))}.
\end{equation}
To fit lattice data for $\chi^Q_2/T^2$ \cite{Borsanyi:2011sw}, we employed the following parametrization for $f_Q (\phi)$
\begin{equation}
\label{eq:fQpar}
f_Q (\phi) = d_1 \, \mathrm{sech}(d_2 \phi) + d_3 \, \mathrm{sech}(d_4 \phi),
\end{equation}
where $d_1 = 0.0193$, $d_2 = -100$, $d_3 = 0.0722$, and $d_4 = 10^{-7}$ (the fourth parameter $d_4$ is introduced for numerical stability and could be taken to be any small number). The overall normalization of the electric coupling $f_Q (\phi)$ is not fixed by fitting lattice data for $\chi^Q_2/T^2$, since it cancels out in Eq. \eqref{eq:susceptibility2}. We have chosen the overall normalization of $f_Q (\phi)$ in Eq. \eqref{eq:fQpar} such as to agree with the order of magnitude of lattice data for the DC electric conductivity in $(2+1)$-flavor QCD \cite{Aarts:2014nba}, as we are going to discuss in the next Section.

\begin{figure}[htp!]
\center
\subfigure[]{\includegraphics[width=0.45\linewidth]{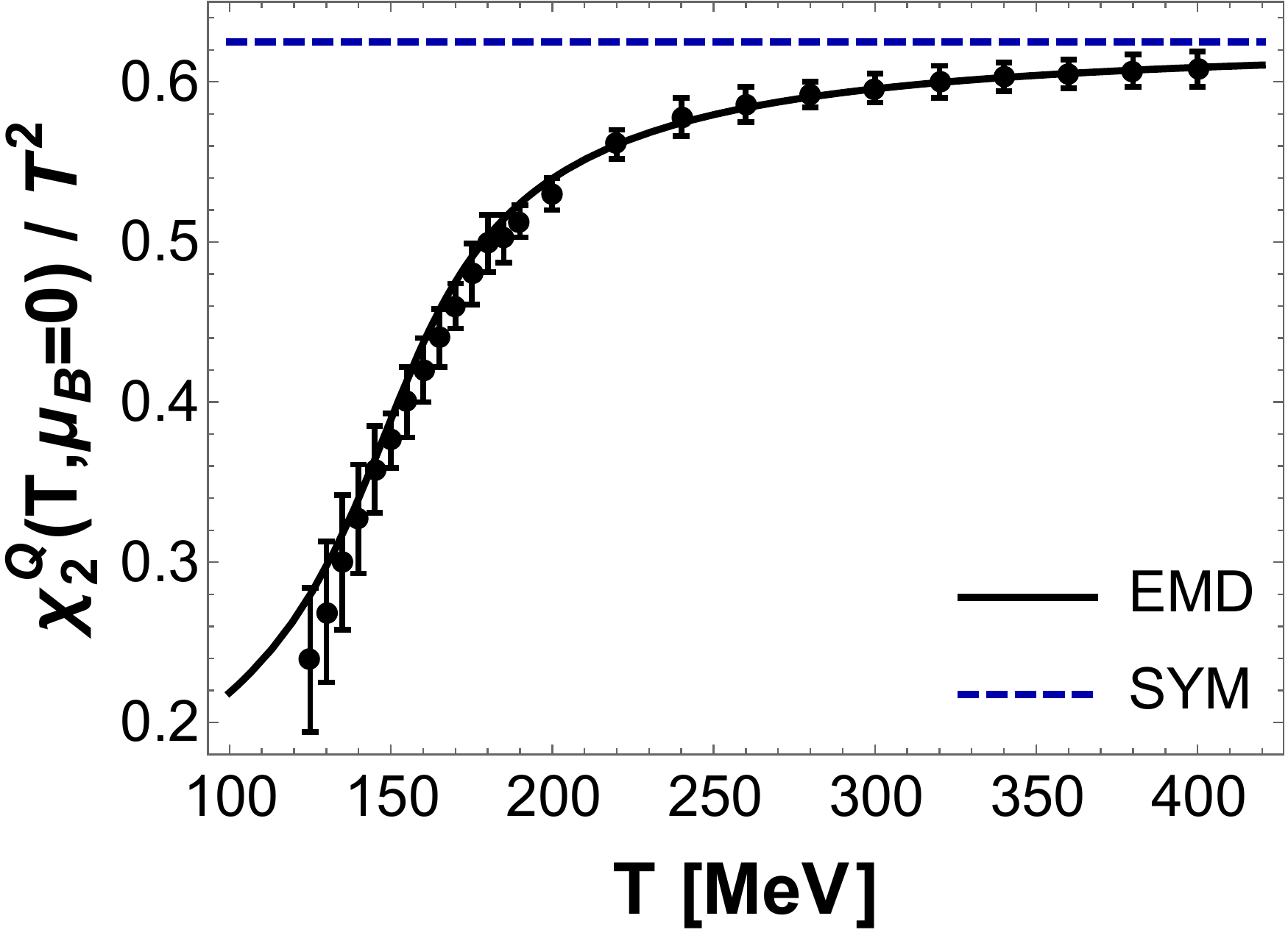}}
\qquad
\subfigure[]{\includegraphics[width=0.45\linewidth]{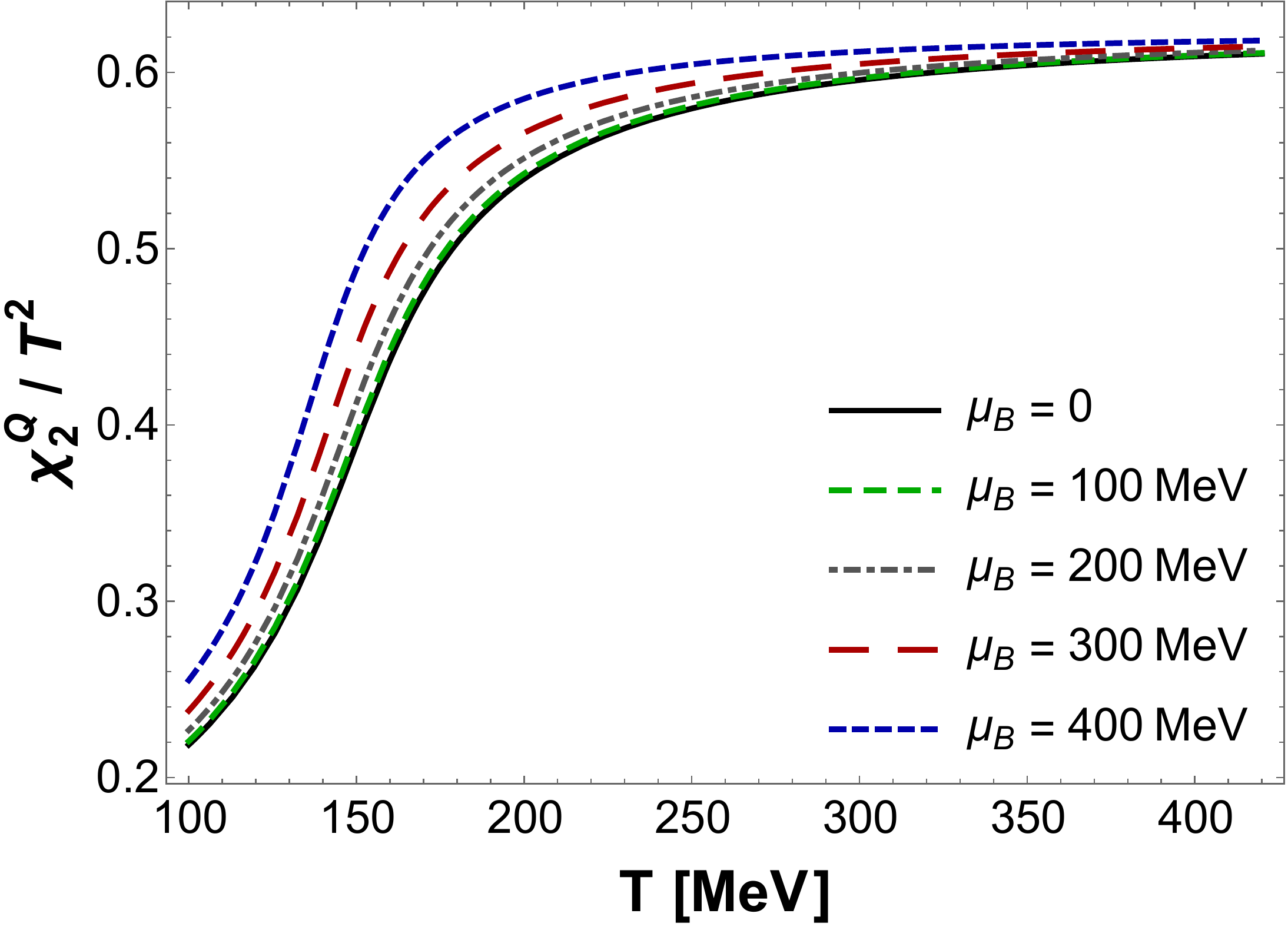}}
\caption{(Color online) Normalized electric charge susceptibility as a function of the temperature $T$ obtained in the present EMD holographic model at: (a) $\mu_B=0$ compared to lattice data for $(2+1)$-flavor QCD with physical quark masses from Ref. \cite{Borsanyi:2011sw} and the $\mathcal{N}=4$ SYM result obtained in Ref. \cite{CaronHuot:2006te}; (b) different fixed values of $\mu_B$.}
\label{fig:electricchi2}
\end{figure}

In Fig. \ref{fig:electricchi2}, we show the holographic results for $\chi^Q_2/T^2$ at (a) zero and (b) nonzero $\mu_B$. One observes that the parametrization given in Eq. \eqref{eq:fQpar} provides a good fit to lattice data at $\mu_B =0$. Moreover, the holographic model predicts that $\chi^Q_2/T^2$ increases with increasing $\mu_B$.

Let us now discuss a possible way to compare the present holographic model result for the electric susceptibility at $\mu_B=0$ with the strongly coupled $\mathcal{N}=4$ SYM result obtained in Ref. \cite{CaronHuot:2006te}, as done in Fig. \ref{fig:electricchi2} (a). In $\mathcal{N}=4$ SYM, $N_c^2=\pi L^3/2G_{5,\,\textrm{SYM}}$ \cite{Gubser:1996de}
and in the present work we take the AdS radius $L$ equals unity, as mentioned before. In the case of the bottom-up EMD action \eqref{eq:EMDaction}, the precise relation between $N_c$ and $G_5$ is unknown. And, in fact, in Eq. \eqref{eq:potential} we have phenomenologically fixed $G_5$ in our EMD model by fitting the $(2+1)$-flavor $N_c=3$ lattice QCD equation of state at $\mu_B=0$. Strictly speaking, in any classical supergravity approximation of the gauge/gravity correspondence, as the one we pursue here, $N_c\rightarrow\infty$. However, as shown in Ref. \cite{Panero:2009tv} for a pure Yang-Mills (YM) plasma, the lattice results for the YM equations of state with $N_c=3,4,5,6$, and 8 are remarkably close to each other, which indicates in practice that, in a sense, at least for some physical observables, ``$N_c\rightarrow\infty$ is not that far from $N_c=3$''. Then, one may consider that, for bottom-up phenomenological applications of the gauge/gravity correspondence, it may be indeed a good idea to fix the free parameters of the action by fitting lattice data for thermodynamic observables. In \cite{CaronHuot:2006te}, it was shown that the electric susceptibility in $\mathcal{N}=4$ SYM is given by
\begin{align}
\frac{\chi_2^{Q,\,\textrm{SYM}}}{T^2}=\frac{N_c^2}{8}.
\label{eq:chiSYM}
\end{align}
For very large values of $T$, corresponding to the UV limit of our EMD holographic model, we have checked that the numerical result for $\chi_2^Q/T^2$ tends to stabilize around the constant value $0.625$. For purposes of comparison, one may adopt a convention where this constant value matches the $\mathcal{N}=4$ SYM result given in Eq. \eqref{eq:chiSYM}, in which case one fixes an ``effective phenomenological holographic value of $N_c^2=5$''. This value was used to plot the $\mathcal{N}=4$ SYM result in Fig. \ref{fig:electricchi2} (a) and it shall be also employed in what follows whenever $N_c^2$ appears explicitly in the holographic calculations.

\subsubsection{DC and AC electric conductivities}
\label{sec:DCandAC}

\begin{figure}[htp!]
\center
\subfigure[]{\includegraphics[width=0.45\linewidth]{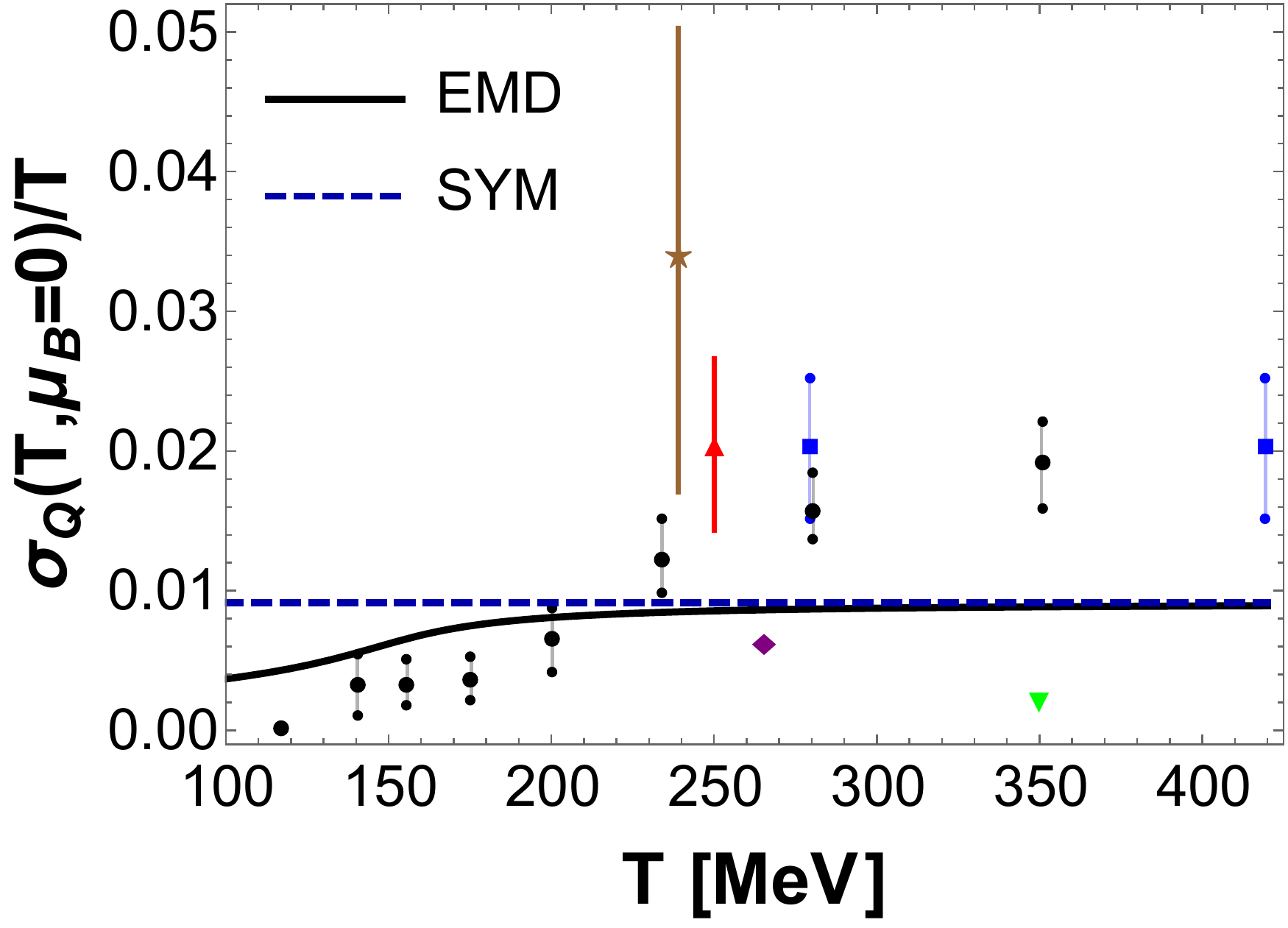}}
\qquad
\subfigure[]{\includegraphics[width=0.45\linewidth]{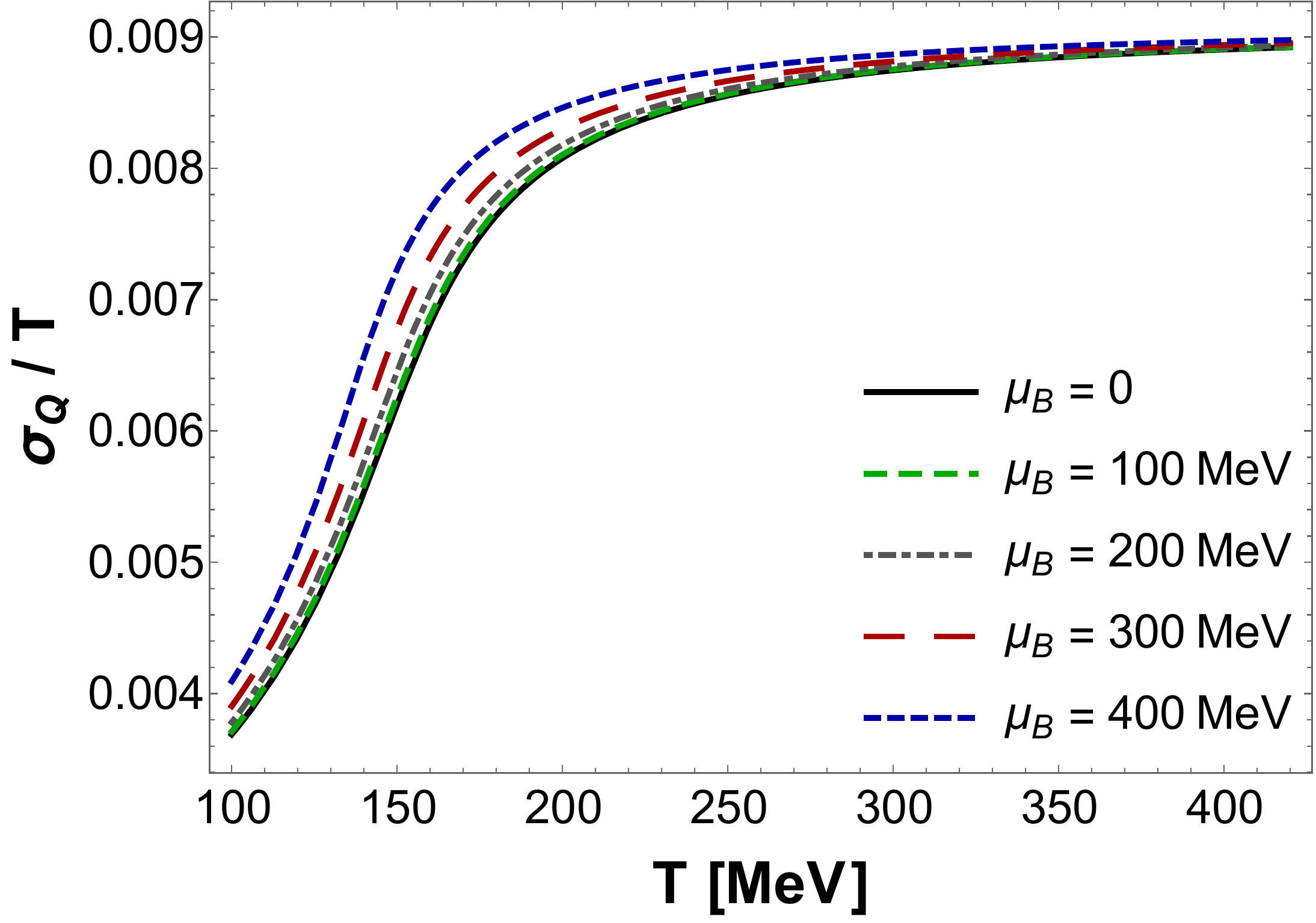}}
\caption{(Color online) Normalized electric DC conductivity as a function of the temperature $T$ obtained in the present EMD holographic model at: (a) $\mu_B=0$ compared to lattice data for $(2+1)$-flavor QCD from Ref. \cite{Aarts:2014nba} and the $\mathcal{N}=4$ SYM result obtained in Ref. \cite{CaronHuot:2006te}; we also show other lattice estimates from Ref.'s \cite{Aarts:2007wj} (blue squares), \cite{Brandt:2012jc} (red triangle), \cite{Buividovich:2010tn} (green inverted triangle), \cite{Burnier:2012ts} (purple diamond), and \cite{Ding:2010ga} (brown star); (b) different fixed values of $\mu_B$.}
\label{fig:sigmaDCT}
\end{figure}

At zero spatial momentum, due to rotational symmetry, it follows that $\sigma^{ij}(\omega)=\sigma(\omega)\delta^{ij}$. The DC electric conductivity $\sigma_Q$ is given by the zero frequency limit of the real part of $\sigma(\omega)$. From the membrane paradigm, one obtains that \cite{Iqbal:2008by}
\begin{align}
\sigma_Q=\lim_{\omega\rightarrow 0}\textrm{Re}[\sigma(\omega)]=\sigma_L(r_H)=\sigma_T(r_H)=\Sigma(r_H).
\label{eq:DCsigma}
\end{align}
We show in Fig. \ref{fig:sigmaDCT} (a) the EMD holographic prediction for the DC electric conductivity at $\mu_B=0$ compared to the latest lattice data available for this observable from Ref. \cite{Aarts:2014nba} (and also some other lattice calculations) and the $\mathcal{N}=4$ SYM result. One notes that the EMD calculation is able to reproduce qualitatively the increase in $\sigma_Q/T$ around the crossover region and also the order of magnitude of this observable as recently obtained in $(2+1)$-flavor lattice QCD simulations \cite{Aarts:2014nba}. In Fig. \ref{fig:sigmaDCT} (b), we display the EMD predictions for $\sigma_Q/T$ at finite values of the baryon chemical potential, from which one notes just a modest rising in the magnitude of the DC electric conductivity with increasing $\mu_B$.

As stated in the previous Section, we have chosen the overall normalization of $f_Q(\phi)$ in Eq. \eqref{eq:fQpar} such as to agree with the order of magnitude of the lattice results for the DC electric conductivity in $(2+1)$-flavor QCD from Ref. \cite{Aarts:2014nba}. The electric conductivity and also the electric diffusion constant and the thermal photon and dilepton production rates are all proportional to the overall normalization of $f_Q(\phi)$, that is, this normalization controls the height of the EMD holographic curves for these observables. We may fix the height of the EMD curve for the DC electric conductivity (at $\mu_B=0$) to lie somewhere in between the lowest and the highest points obtained in Ref. \cite{Aarts:2014nba}. It turns out that this criterion may be satisfied by imposing that in the ultraviolet the EMD result for $\sigma_Q(T,\mu_B=0)/T$ tends to the corresponding $\mathcal{N}=4$ SYM result obtained in Ref. \cite{CaronHuot:2006te},
\begin{align}
\frac{\sigma_{Q,\,\textrm{SYM}}}{T}=\frac{e^2N_c^2}{16\pi}=\frac{\alpha_{\textrm{QED}}N_c^2}{4}= \frac{(1/137)\times 5}{4}\approx 0.0091,
\label{eq:condconf}
\end{align}
where we used the ``effective phenomenological holographic value of $N_c^2=5$'' discussed in the last Section. We must remark, however, that $\mathcal{N}=4$ SYM theory and $(2+1)$-flavor QCD are very different theories and that our actual guide to fix the overall normalization of $f_Q(\phi)$ is the lattice data from Ref. \cite{Aarts:2014nba}, instead of Eq. \eqref{eq:condconf}. In fact, we could have as well just slightly modified the overall normalization of $f_Q(\phi)$ by performing some small shift on the height of the EMD curve for $\sigma_Q(T,\mu_B=0)/T$ as long as it stayed in between the lowest and the highest points of Ref. \cite{Aarts:2014nba}. Such a small shift would imply the very same small shift in the height of the electric diffusion constant and the thermal photon and dilepton production rates, but none of our conclusions in the present work would be modified.

\begin{figure}[htp!]
\center
\subfigure[]{\includegraphics[width=0.46\linewidth]{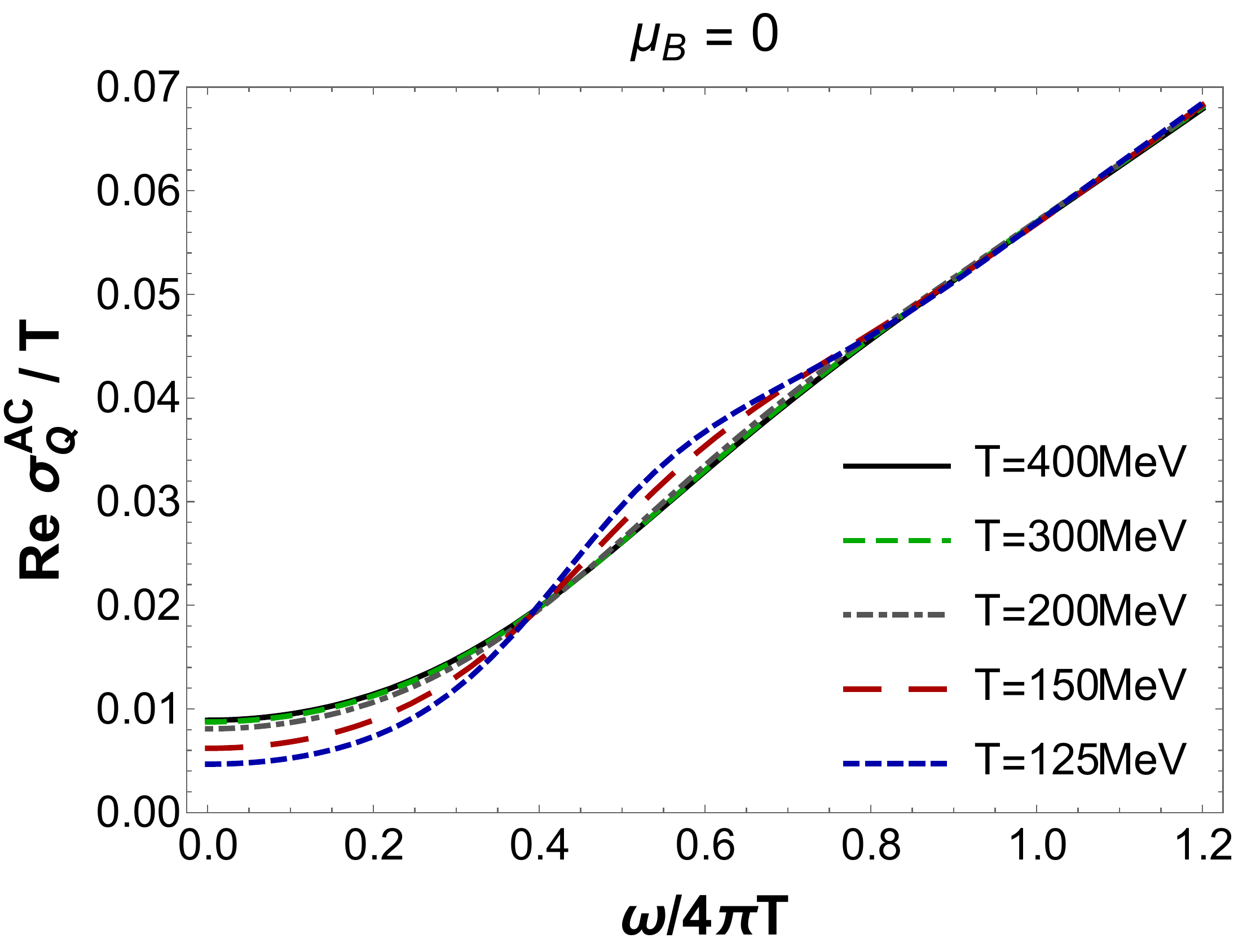}}
\qquad
\subfigure[]{\includegraphics[width=0.46\linewidth]{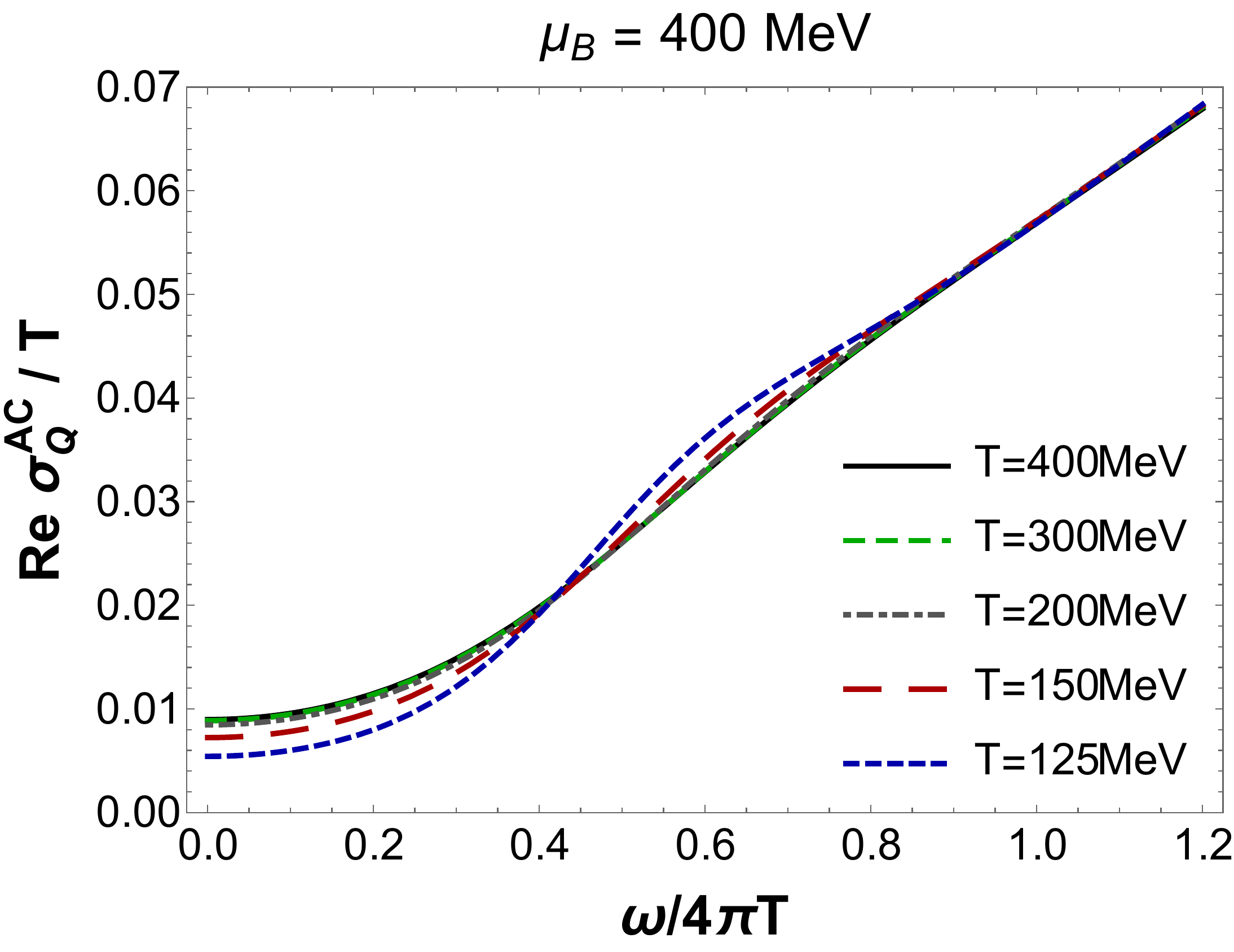}}
\caption{(Color online) Real part of the normalized AC electric conductivity as a function of frequency for several fixed values of the temperature $T$ at (a) $\mu_B=0$ and (b) $\mu_B = 400 \, \mathrm{MeV}$.}
\label{fig:sigmaAC}
\end{figure}

The AC electric conductivity $\sigma_Q^{\textrm{AC}}(\omega)$ is obtained by numerically solving either of the flow equations \eqref{eq:flow1} or \eqref{eq:flow2} at zero spatial momentum (both flow equations coincide at $k=0$). In Fig. \ref{fig:sigmaAC}, we show the EMD holographic results for the real part of $\sigma_Q^{\textrm{AC}}(\omega)/T$ as a function of frequency, temperature, and baryon chemical potential. One sees that increasing the temperature and/or the baryon chemical potential dampens the oscillations observed in the AC conductivity as a function of frequency.

\subsubsection{Electric diffusion constant}
\label{sec:diffusion}

\begin{figure}[htp!]
\center
\subfigure[]{\includegraphics[width=0.45\linewidth]{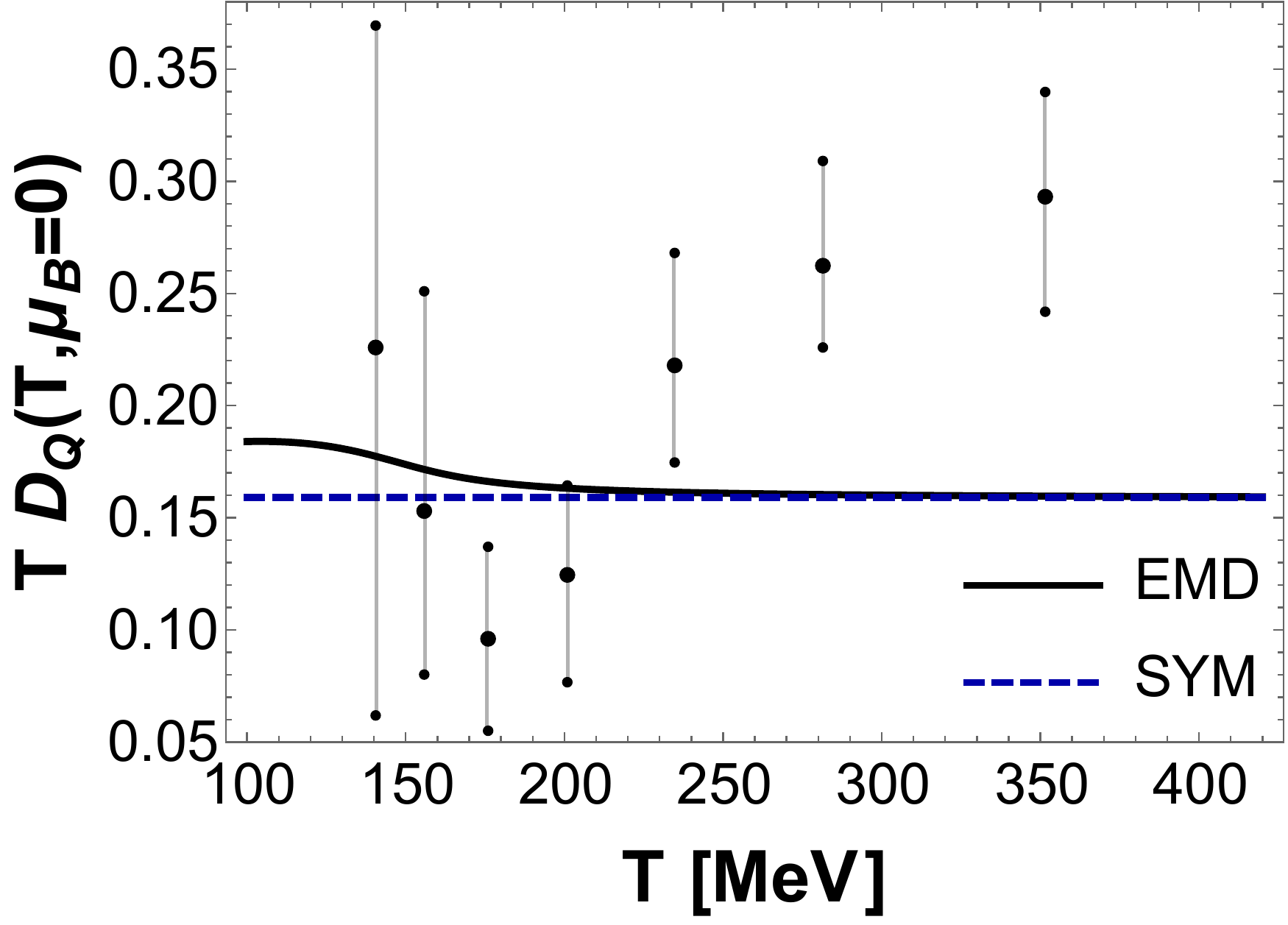}}
\qquad
\subfigure[]{\includegraphics[width=0.45\linewidth]{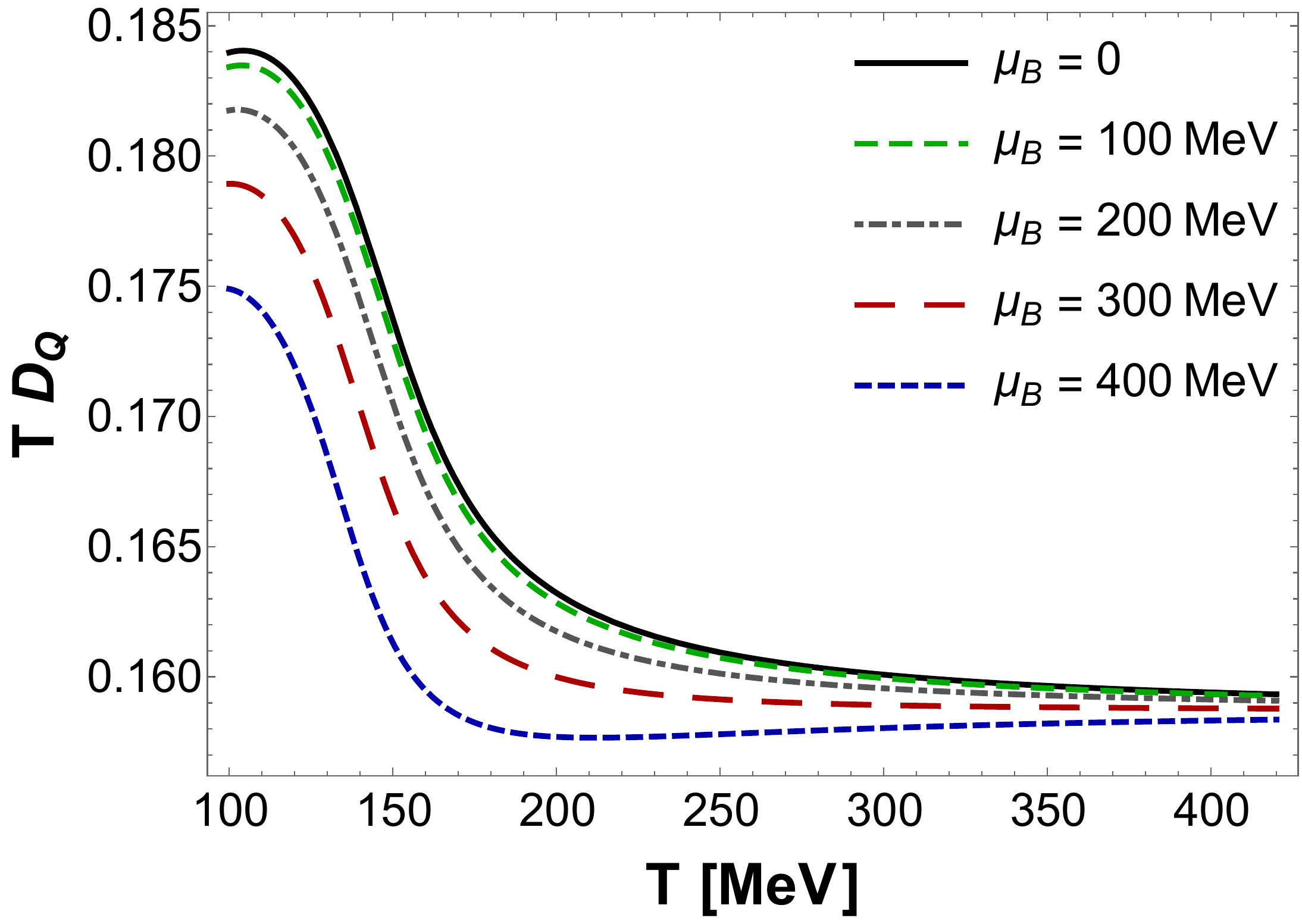}}
\caption{(Color online) Normalized electric charge diffusion constant as a function of the temperature $T$ obtained in the present EMD holographic model at: (a) $\mu_B=0$ compared to lattice data for $(2+1)$-flavor QCD from Ref. \cite{Aarts:2014nba} and the $\mathcal{N}=4$ SYM result obtained in Ref. \cite{Policastro:2002se}; (b) different fixed values of $\mu_B$.}
\label{fig:DT}
\end{figure}

It is possible to show from the flow equations that for any charged or uncharged black brane the Nernst-Einstein's relation for the diffusion constant $D_Q$ holds \cite{Iqbal:2008by}
\begin{equation}
D_Q = \frac{\sigma_Q}{e^2\chi_2^Q}.
\end{equation}
In Fig. \ref{fig:DT} (a), we show the EMD holographic prediction for the normalized diffusion constant $TD_Q$ at $\mu_B=0$ compared to recent lattice data from Ref. \cite{Aarts:2014nba} and also the $\mathcal{N}=4$ SYM result $TD_{Q,\,\textrm{SYM}}=1/2\pi$ originally obtained in Ref. \cite{Policastro:2002se}. One observes that the EMD model quantitatively captures the decrease in $TD_Q$ from lower temperatures up to the crossover region as seen in lattice simulations, although it misses the lattice behavior for higher temperatures. We also show in Fig. \ref{fig:DT} (b) the EMD predictions for $TD_Q$ at finite values of the baryon chemical potential, from which one sees a suppression of electric diffusion transport with increasing $\mu_B$. The same conclusion of suppression of diffusive transport at nonzero baryon chemical potential was shown to hold also for the diffusion of baryon charge in Ref. \cite{Rougemont:2015ona}. Therefore, from the present phenomenological EMD holographic model one extracts the general prediction of an overall suppression of diffusive transport as one moves toward increasing values of the baryon density.

\section{Thermal photon and dilepton production}
\label{sec:spectra}

\subsection{Spectral function}
\label{sec:spectralfun}

We now proceed to the main focus of the present work: the evaluation of thermal photon and dilepton production rates in a strongly coupled QGP at both zero and nonzero baryon chemical potential. To accomplish this goal, we need first to obtain an explicit expression for the trace of the spectral function \eqref{eq:trace} in terms of the boundary values of $\sigma_L$ and $\sigma_T$ defined in Eq. \eqref{eq:defsigmas}.

As done before, we take $\tilde{k}^\mu=(\tilde{\omega},0,0,\tilde{k})$, and by using Eq. \eqref{eq:Kubo}, the Ward's identity $\tilde{k}_\mu G^{\mu\nu}_R(\tilde{k}^\mu)=0$ implies the following system of equations
\begin{align}
-\tilde{\omega}\sigma^{tt}+\tilde{k}\sigma^{zt}&=0,\nonumber\\
-\tilde{\omega}\sigma^{tz}+\tilde{k}\sigma^{zz}&=0,
\end{align}
which in turn gives the constraint
\begin{align}
\sigma^t_t=-\frac{\tilde{k}^2}{\tilde{\omega}^2}\sigma_L,
\label{eq:WardId}
\end{align}
where $\sigma_L=\sigma_z^z$. Using that $\sigma_T=\sigma_x^x=\sigma_y^y$ and Eq.'s \eqref{eq:trace} and \eqref{eq:WardId}, it follows that the trace of the spectral function reads as follows
\begin{align}
\chi^\mu_\mu(\tilde{k}^\mu)=2\tilde{\omega}\left(2\textrm{Re}[\sigma_T(\tilde{k}^\mu,\tilde{r}\rightarrow\infty)]+ \left(1-\frac{\tilde{k}^2}{\tilde{\omega}^2}\right)\textrm{Re}[\sigma_L(\tilde{k}^\mu,\tilde{r}\rightarrow\infty)]\right).
\label{eq:GeneralTrace}
\end{align}
Using Eq. \eqref{eq:scalingmomentum}, one may translate Eq. \eqref{eq:GeneralTrace} to the numerical coordinates as below
\begin{align}
\chi^\mu_\mu(k^\mu)=8\pi T\omega\left(2\textrm{Re}[\sigma_T(k^\mu,r\rightarrow\infty)]+ \left(1-\frac{k^2}{\omega^2}h_0^{\textrm{far}}\right)\textrm{Re}[\sigma_L(k^\mu,r\rightarrow\infty)]\right).
\label{eq:GeneralTraceNum}
\end{align}

\subsection{Thermal photon production rate}
\label{sec:photonproduction}

At leading order in the electromagnetic coupling constant $\alpha_{\textrm{QED}}$ but at all orders in the strong coupling constant $\alpha_s$, the normalized thermal photon production rate is given by \cite{CaronHuot:2006te,Mamo:2013efa,Yang:2015bva}
\begin{equation}
\frac{d\hat{\Gamma}_{\gamma}}{d\tilde{\omega}}\equiv\frac{1}{\alpha_{\textrm{QED}}N_c^2T^3} \frac{d\Gamma_{\gamma}}{d\tilde{\omega}}=\frac{\bar{Q}_\gamma}{\alpha_{\textrm{QED}}N_c^2T^3} \chi_\mu^\mu(\tilde{k}^\mu)\biggr|_{\tilde{k}^\mu=(\tilde{\omega},0,0,\tilde{\omega})},
\label{eq:photonrate0}
\end{equation}
where the trace of the spectral function is evaluated at light-like momentum $\tilde{k}=\tilde{\omega}$ and
\begin{align}
\bar{Q}_\gamma=\frac{\alpha_{\textrm{QED}}T}{\pi}\frac{(\tilde{\omega}/T)}{e^{\tilde{\omega}/T}-1}= \frac{4\alpha_{\textrm{QED}}\,\omega T}{e^{4\pi\omega}-1},
\end{align}
where in the last step we used Eq. \eqref{eq:scalingmomentum}. Again, by using Eq. \eqref{eq:scalingmomentum}, the light-like momentum condition $\tilde{k}=\tilde{\omega}$ reads as $k=\omega/\sqrt{h_0^{\textrm{far}}}$ in the numerical coordinates, such that Eq. \eqref{eq:photonrate0} is translated to the numerical coordinates as below
\begin{equation}
\frac{d\hat{\Gamma}_{\gamma}}{d\tilde{\omega}}=\frac{4\omega}{N_c^2T^2(e^{4\pi\omega}-1)} \chi_\mu^\mu(k^\mu)\biggr|_{k^\mu=(\omega,0,0,\omega/\sqrt{h_0^{\textrm{far}}})}.
\label{eq:photonrate}
\end{equation}
Note that by applying the light-like momentum condition $k=\omega/\sqrt{h_0^{\textrm{far}}}$ in Eq. \eqref{eq:GeneralTraceNum} for the trace of the spectral function, one concludes that only the transverse conductivity contributes to the photon production rate through
\begin{equation}
\chi_\mu^\mu(k^\mu)\biggr|_{k^\mu=(\omega,0,0,\omega/\sqrt{h_0^{\textrm{far}}})}=16\pi T\omega\textrm{Re}[ \sigma_T(k^\mu,r\rightarrow\infty)]\biggr|_{k^\mu=(\omega,0,0,\omega/\sqrt{h_0^{\textrm{far}}})}.
\end{equation}

\begin{figure}[htp!]
\center
\subfigure[]{\includegraphics[width=0.46\linewidth]{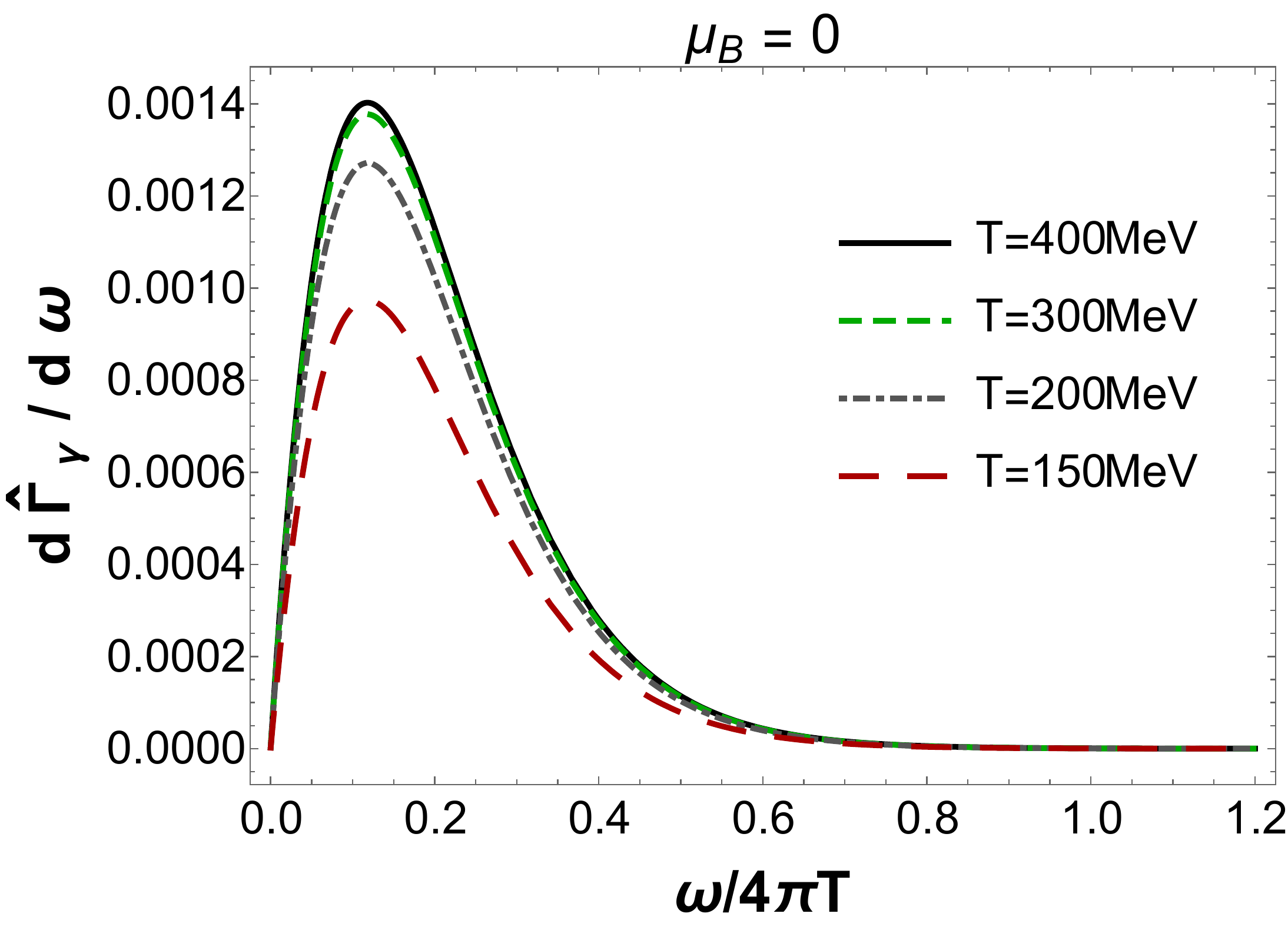}}
\qquad
\subfigure[]{\includegraphics[width=0.46\linewidth]{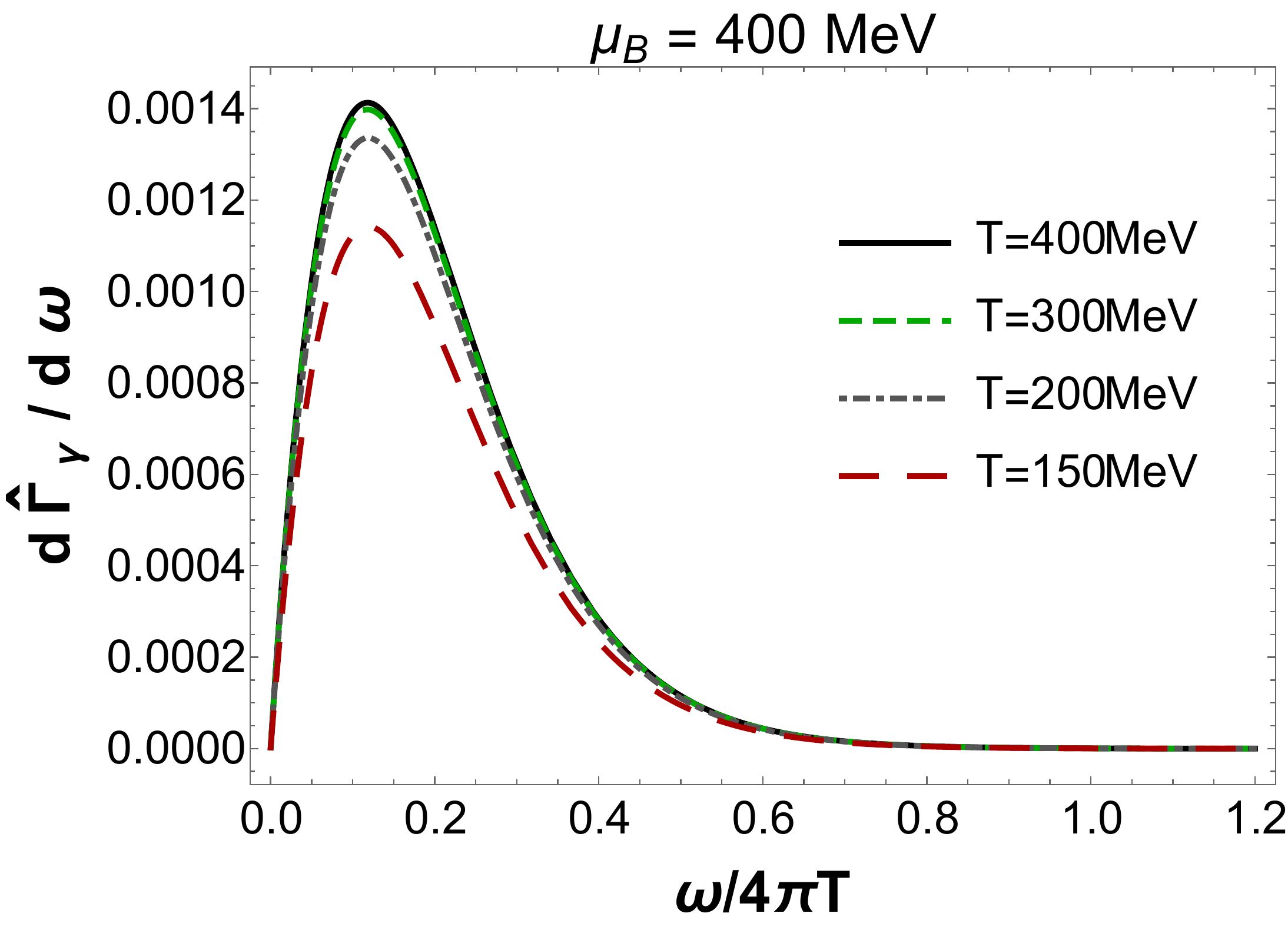}}
\caption{(Color online) Normalized thermal photon production rate as a function of frequency for several fixed values of the temperature $T$ at (a) $\mu_B=0$ and (b) $\mu_B = 400 \, \mathrm{MeV}$.}
\label{fig:thermalphoton}
\end{figure}

\begin{figure}[htp!]
\center
\begin{centering}
\includegraphics[scale=0.52]{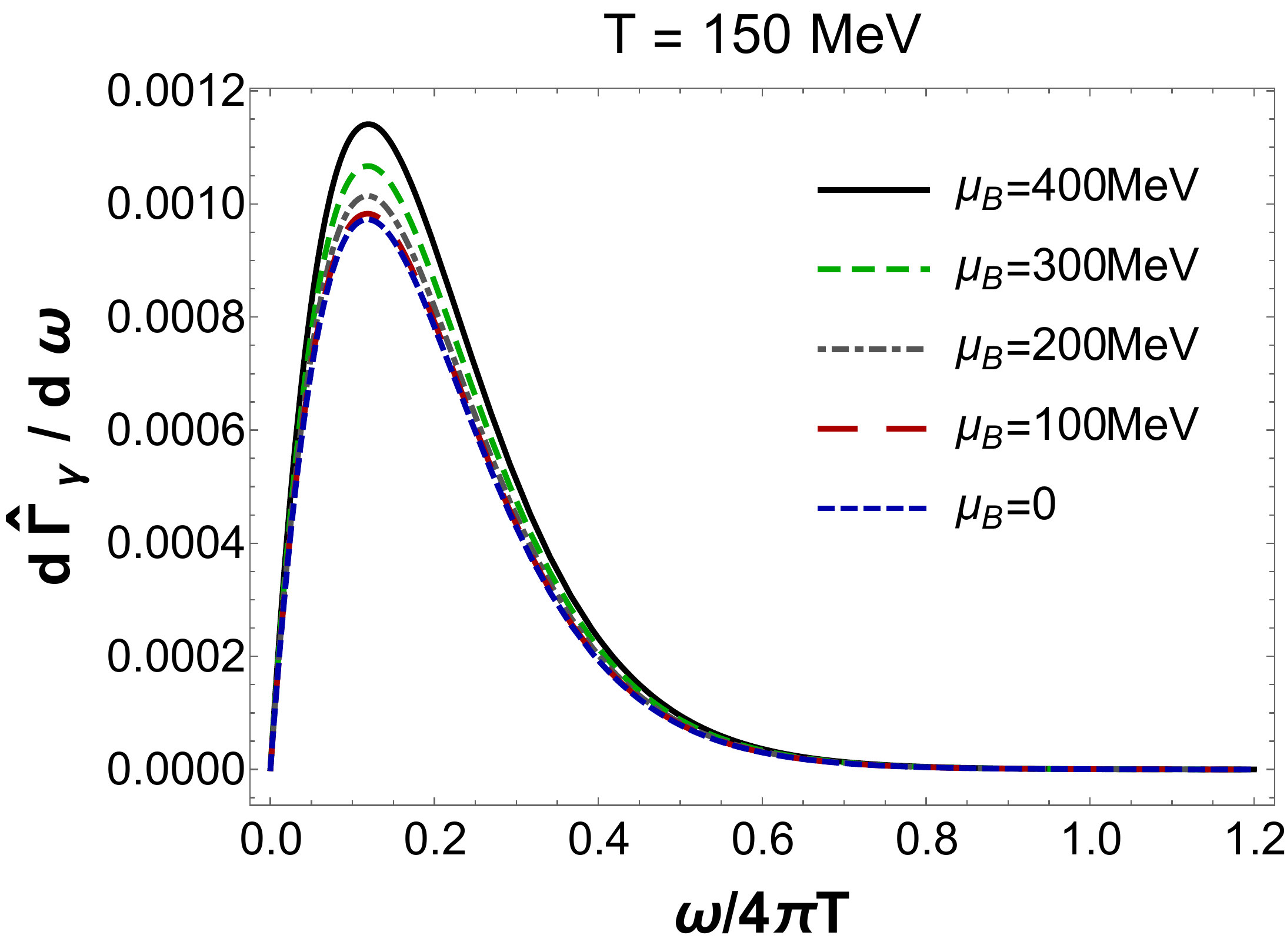}
\par\end{centering}
\caption{(Color online) Normalized thermal photon production rate as a function of frequency for several fixed values of the baryon chemical potential $\mu_B$ at $T = 150 \, \mathrm{MeV}$.}
\label{fig:thermalphoton2}
\end{figure}

In Fig. \ref{fig:thermalphoton}, we present the EMD holographic predictions for the normalized thermal photon production rate $d\hat{\Gamma}_{\gamma}/d\omega$ given by Eq. \eqref{eq:photonrate0} (or also Eq. \eqref{eq:photonrate}) at zero and nozero values of the baryon chemical potential. We see that increasing the temperature $T$ enhances the peak of the photon production rate; the same effect, albeit more discreetly, is seen when one increases the baryon chemical potential $\mu_B$. This last effect can be seen more clearly in Fig. \ref{fig:thermalphoton2}, where we fixed $T = 150 \, \mathrm{MeV}$ (the lower the temperature the higher the effect of a non-vanishing baryon chemical potential). It is interesting to note that, besides the enhancement observed in the EMD model for the peak of the thermal photon production rate as one increases the temperature and/or the baryon chemical potential, an enhancement in the peak of the $\mathcal{N} = 4$ SYM thermal photon production rate including the leading order corrections in the 't Hooft coupling $\lambda_t$ was obtained in Ref.'s \cite{Hassanain:2011ce,Hassanain:2012uj} as one of the main effects of decreasing the value of $\lambda_t$.

It is also instructive to compare our results for the normalized thermal photon production rate at different fixed values of $T$ and $\mu_B=0$ shown in Fig. \ref{fig:thermalphoton} (a) with the corresponding results obtained in Ref. \cite{Yang:2015bva} (see Fig. 3 of that work). One observes that both predictions agree qualitatively, although the results found in Ref. \cite{Yang:2015bva} are one order of magnitude above our results. Such a quantitative discrepancy is entirely due to the different overall normalizations chosen for the Maxwell-Dilaton electric coupling function $f_Q(\phi)$ in these two works. While in the present work we choose to fix the overall normalization of $f_Q(\phi)$ by matching the UV asymptotic behavior at large $T$ for the ratio $\sigma_Q/T$ with the corresponding $\mathcal{N} = 4$ SYM result (as discussed before), in Ref. \cite{Yang:2015bva} it was used instead a different normalization, where the UV asymptotic behavior of the normalized thermal photon production rate itself was matched with the corresponding $\mathcal{N} = 4$ SYM result obtained in Ref. \cite{CaronHuot:2006te}. The normalization we have chosen in the present work gives a reasonable agreement with the latest lattice data \cite{Aarts:2014nba} for the DC electric conductivity and the electric diffusion constant around the crossover transition, as shown in Sections \ref{sec:DCandAC} and \ref{sec:diffusion}, respectively. By adopting the normalization used in Ref. \cite{Yang:2015bva}, the holographic results obtained for the electric charge transport sector are one order of magnitude above the lattice data of Ref. \cite{Aarts:2014nba}. However, one must keep in mind the fact that, as displayed in Fig. 6 of Ref. \cite{Greif:2014oia}, there is currently no consensus in the literature regarding the order of magnitude of $\sigma_Q/T$ in QCD, since it may drastically vary depending on different models and calculations and there are discrepancies even among different lattice collaborations. Therefore, although the qualitative behaviors of the thermal photon production rate at $\mu_B=0$ obtained here and in Ref. \cite{Yang:2015bva} are exactly the same, the order of magnitude of this physical observable remains as an open question which deserves further investigation. We remark, however, that if one takes as a guide the latest lattice data \cite{Aarts:2014nba} for $\sigma_Q/T$, our holographic results constraint the order of magnitude of the thermal photon production rate to be as low as shown in Fig. \ref{fig:thermalphoton}. This is roughly one order of magnitude below the perturbative QCD result obtained at $\alpha_s=0.2$ in Ref. \cite{Arnold:2001ms}, and it would lie two orders of magnitude below estimates obtained by employing the values of $\sigma_Q/T$ derived by using the Boltzmann Approach to Multi-Parton Scatterings (BAMPS) \cite{Greif:2014oia} and Parton Hadron String Dynamics (PHSD) \cite{Cassing:2013iz}. The suppression in the magnitude of the thermal photon spectra obtained here when compared to perturbative calculations is consistent with what seems to be a general trend regarding the behavior of non-equilibrium observables in strongly coupled systems, like the QGP around the crossover transition. For instance, the ratio of the shear viscosity over the entropy density obtained by fitting heavy ion experimental data for anisotropic flow coefficients $v_n$ using hydrodynamics gives $\eta/s=0.2$ \cite{Gale:2012rq}, which is of the same order of magnitude of the small value $\eta/s=1/4\pi$ found in a broad class of holographic duals \cite{Kovtun:2004de,Buchel:2003tz} which includes our model and at least one order of magnitude below perturbative QCD calculations \cite{Arnold:2003zc,Fuini:2010xz}. Another example is the suppression of roughly one order of magnitude in the baryon conductivity found in Ref. \cite{Rougemont:2015ona} when compared to the kinetic theory calculations of Ref. \cite{Jaiswal:2015mxa}.

\subsection{Thermal dilepton production rate}
\label{sec:dileptonproduction}

The thermal dilepton production rate is given by \cite{CaronHuot:2006te,Mamo:2013efa,Jahnke:2013rca}
\begin{equation}
\frac{d\Gamma_{l\bar{l}}}{d^4\tilde{k}} = Q_{l\bar{l}}\, \chi^{\mu}_{\mu}(\tilde{k}^\mu)\biggr|_{\tilde{k}^\mu=(\tilde{\omega},0,0,\tilde{k}<\tilde{\omega})},
\end{equation}
where the trace of the spectral function is evaluated at time-like momentum $-\tilde{\omega}^2+\tilde{k}^2 = -M^2 < 0$, with $M$ being the invariant dilepton mass and
\begin{equation}
Q_{l\bar{l}} = \frac{1}{(2\pi)^4} \frac{e^2 e_l^2}{6 \pi |\tilde{k}_\mu^2|^{5/2}} \theta (\tilde{\omega}) \theta( -\tilde{k}_\mu^2 - 4 m_l^2) \sqrt{-\tilde{k}_\mu^2 - 4m_l^2}\, (-\tilde{k}_\mu^2 + 2m_l^2) \frac{1}{e^{\tilde{\omega}/T}-1},
\label{eq:phasefactor}
\end{equation}
where $m_l$ and $e_l$ are the lepton mass and charge, respectively. By following \cite{Mamo:2013efa}, we define the following normalized thermal dilepton production rate
\begin{align}
\frac{d\hat{\Gamma}_{l\bar{l}}}{d^4K}\equiv\frac{1}{\bar{Q}_{l\bar{l}}}\frac{d\Gamma_{l\bar{l}}}{d^4\tilde{k}}= \frac{16\pi}{N_c^2T^2(e^{4\pi\omega}-1)} \chi^{\mu}_{\mu}(k^\mu)\biggr|_{k^\mu=(\omega,0,0,k<\omega/\sqrt{h_0^{\textrm{far}}})},
\label{eq:dileptonproduction}
\end{align}
which is already written in the numerical coordinates. We have also defined
\begin{align}
\bar{Q}_{l\bar{l}}\equiv Q_{l\bar{l}}\,T(e^{\tilde{\omega}/T}-1)\,\frac{\sigma_{Q,\,\textrm{SYM}}}{e^2} =Q_{l\bar{l}}(e^{4\pi\omega}-1)\,\frac{N_c^2T^2}{16\pi},
\end{align}
where in the last step we made use of Eq.'s \eqref{eq:scalingmomentum} and \eqref{eq:condconf}. Note also that for the thermal dilepton production, due to the time-like momentum condition $k<\omega/\sqrt{h_0^{\textrm{far}}}$, both the longitudinal and transverse conductivities contribute to the trace of the spectral function in Eq. \eqref{eq:GeneralTraceNum}.

\begin{figure}[htp!]
\center
\subfigure[]{\includegraphics[width=0.46\linewidth]{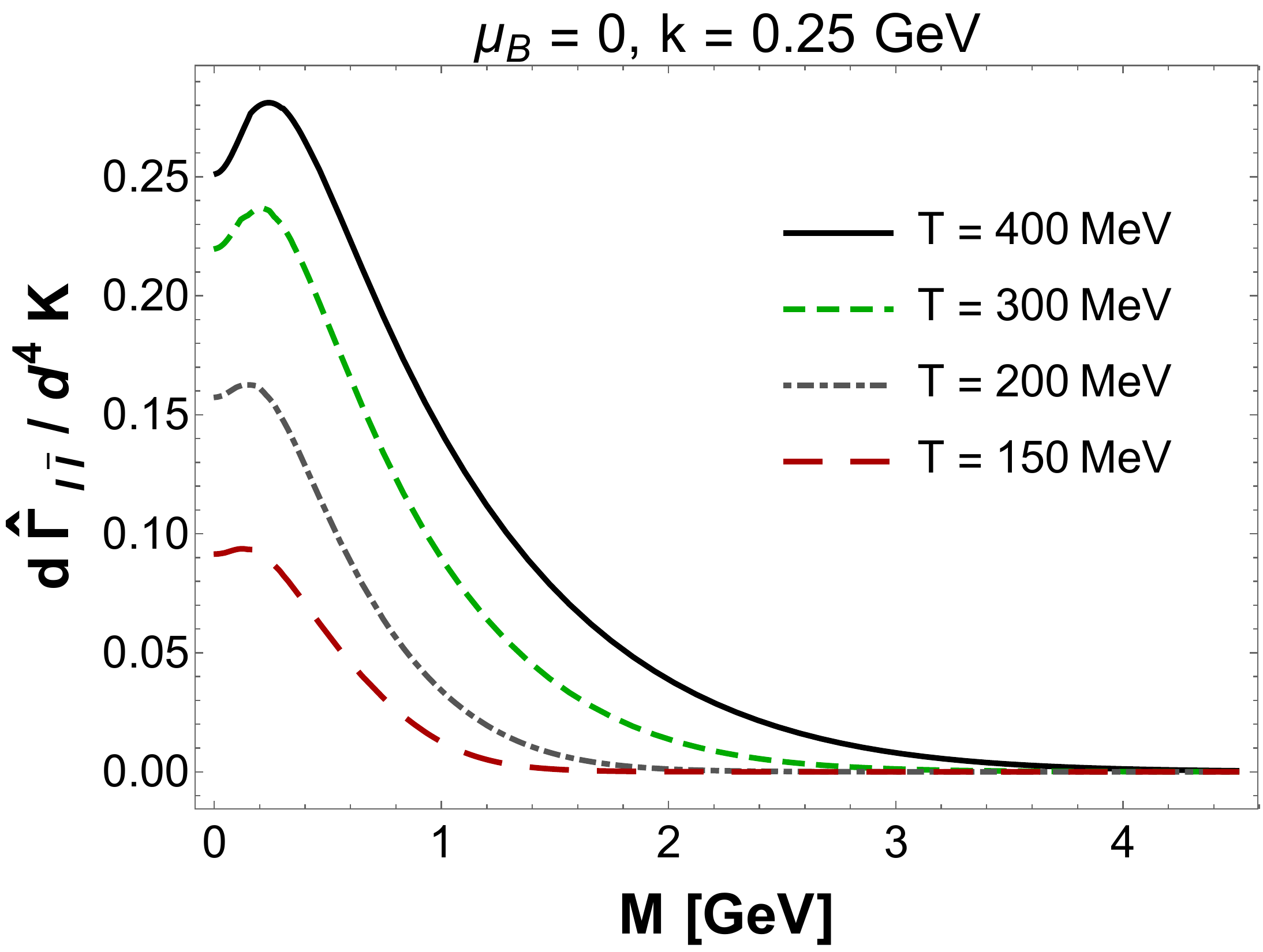}}
\qquad
\subfigure[]{\includegraphics[width=0.46\linewidth]{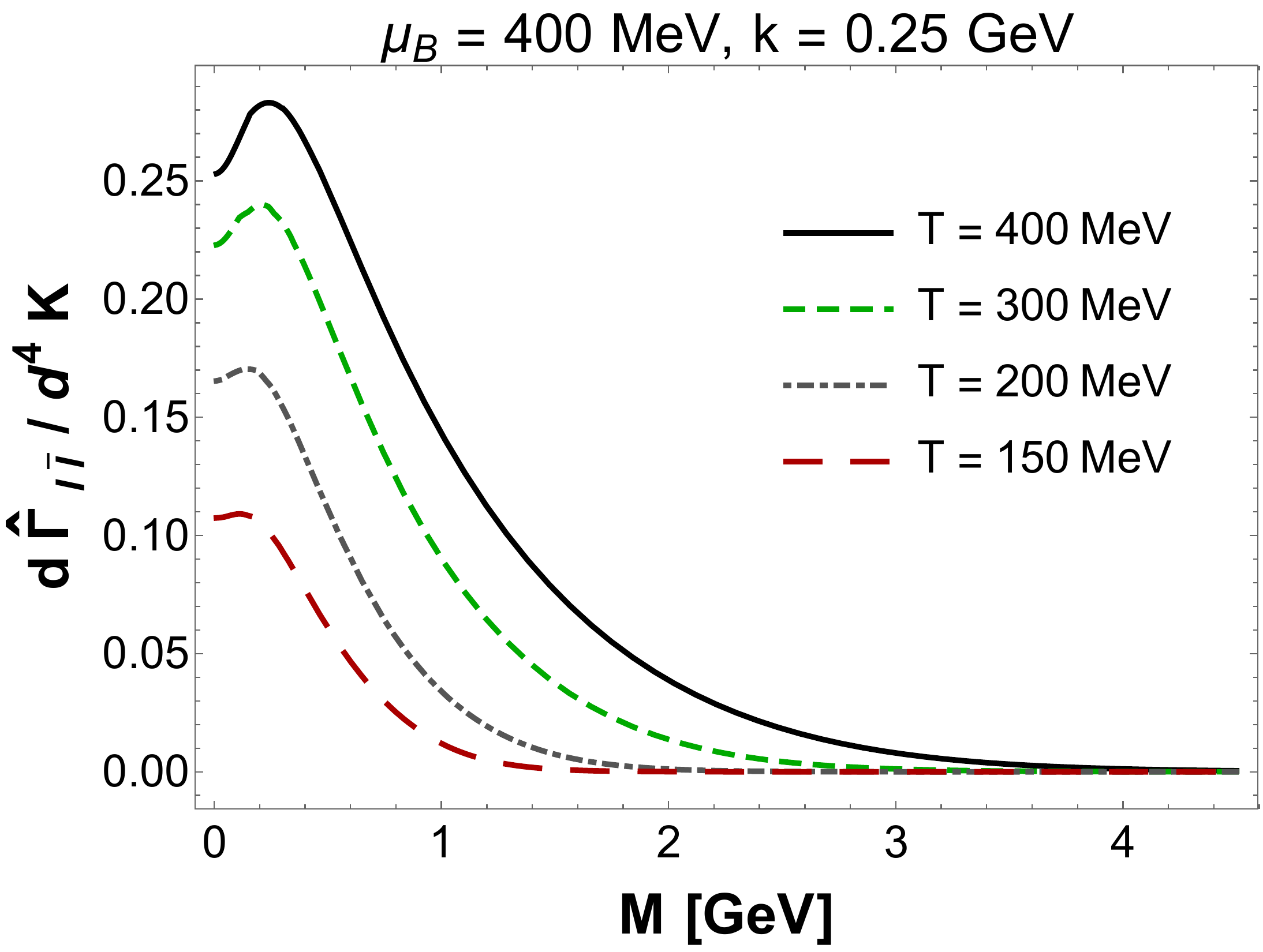}}
\qquad
\subfigure[]{\includegraphics[width=0.46\linewidth]{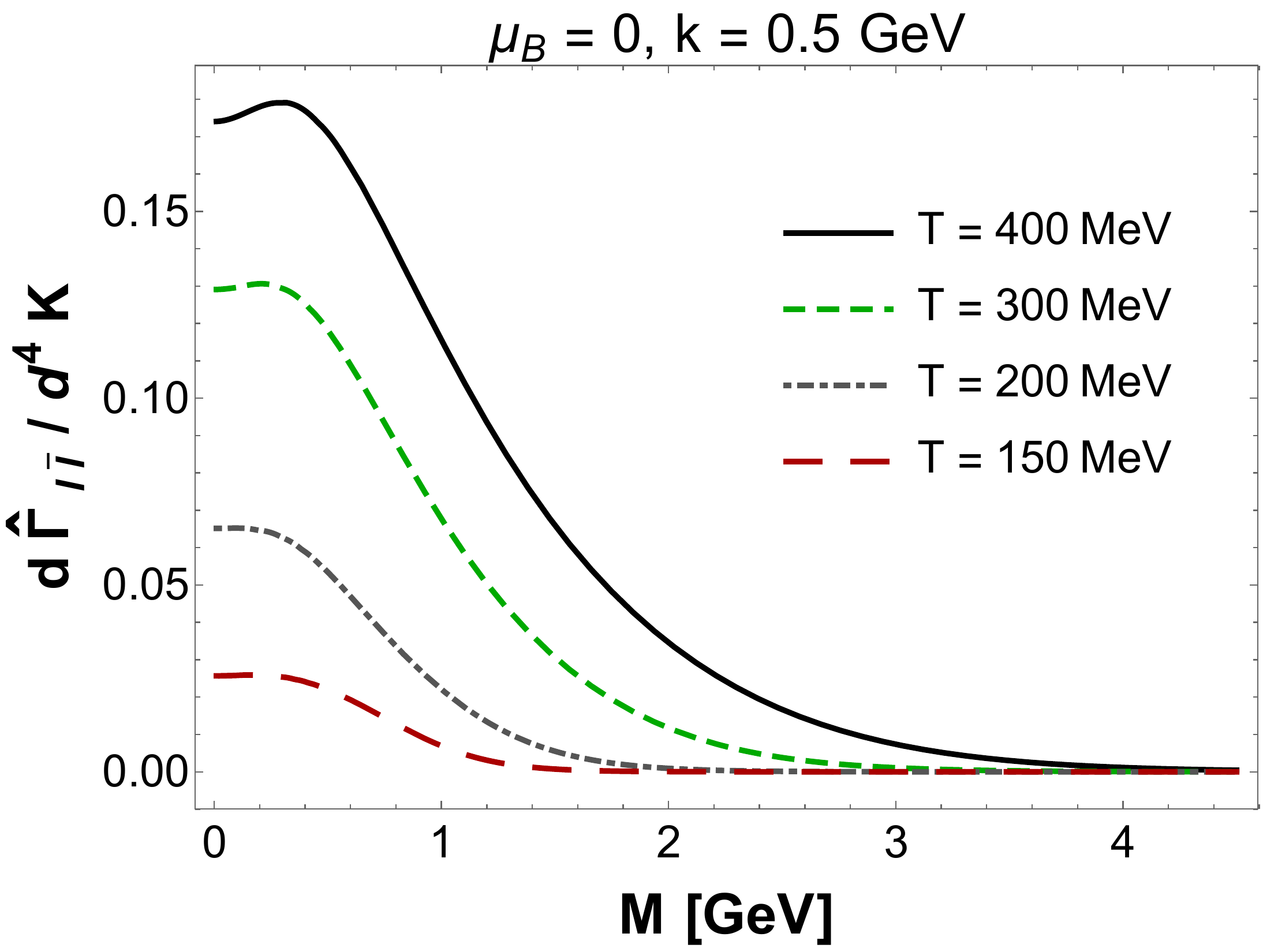}}
\qquad
\subfigure[]{\includegraphics[width=0.46\linewidth]{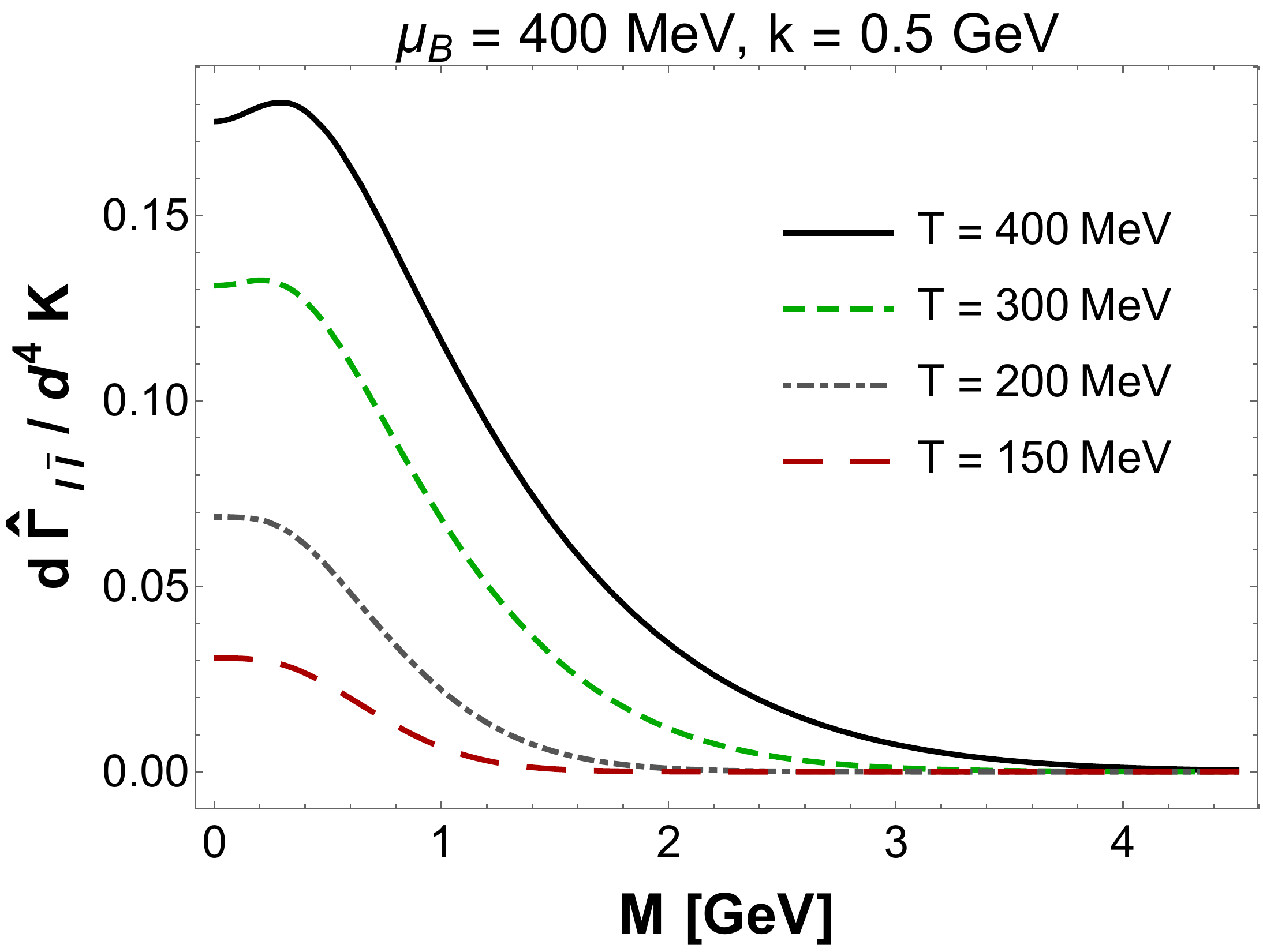}}
\qquad
\subfigure[]{\includegraphics[width=0.46\linewidth]{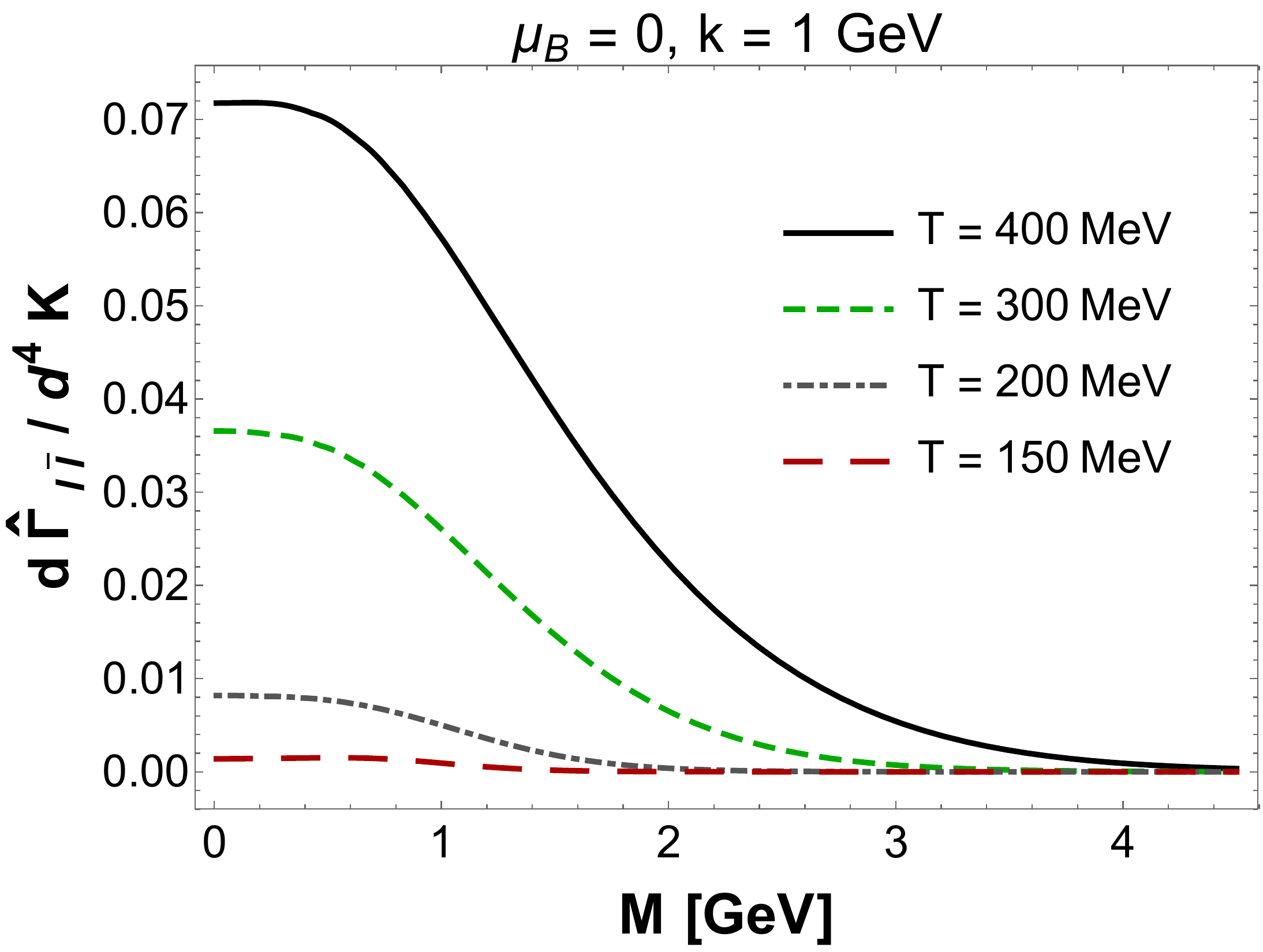}}
\qquad
\subfigure[]{\includegraphics[width=0.46\linewidth]{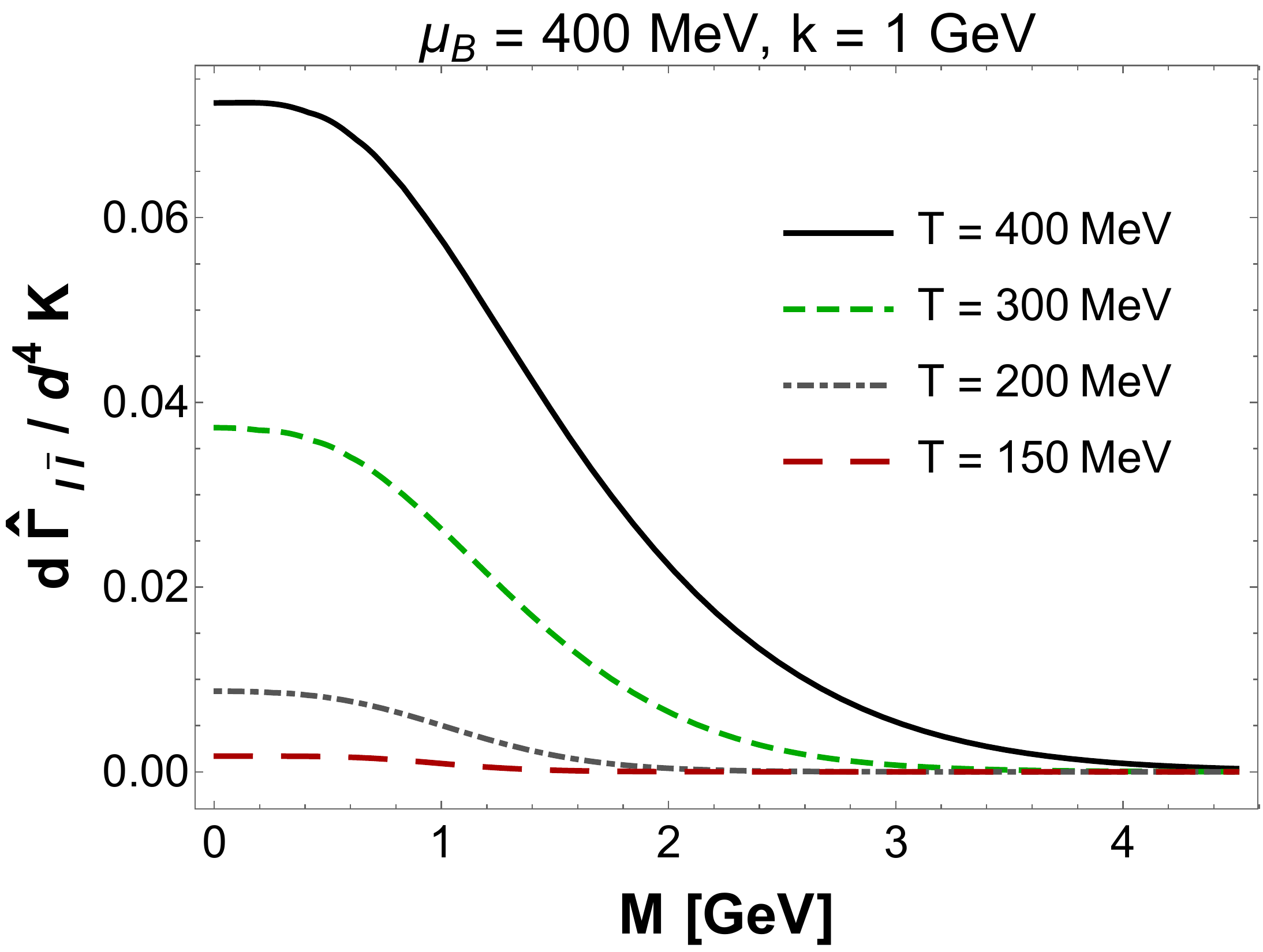}}
\caption{(Color online) Normalized thermal dilepton production rate as a function of the invariant mass $M$ of the dilepton pair for several values of the temperature $T$ and momentum $k$. On the left panel, we show the results at $\mu_B = 0$ and (a) $k = 0.25 \, \mathrm{GeV}$, (c) $k = 0.5 \, \mathrm{GeV}$, and (e) $k = 1 \, \mathrm{GeV}$. On the right panel, the display results at $\mu_B = 400 \, \mathrm{MeV}$ and (b) $k = 0.25 \, \mathrm{GeV}$, (d) $k = 0.5 \, \mathrm{GeV}$, and (f) $k = 1 \, \mathrm{GeV}$.}
\label{fig:thermaldilepton}
\end{figure}

\begin{figure}[htp!]
\center
\subfigure[]{\includegraphics[width=0.46\linewidth]{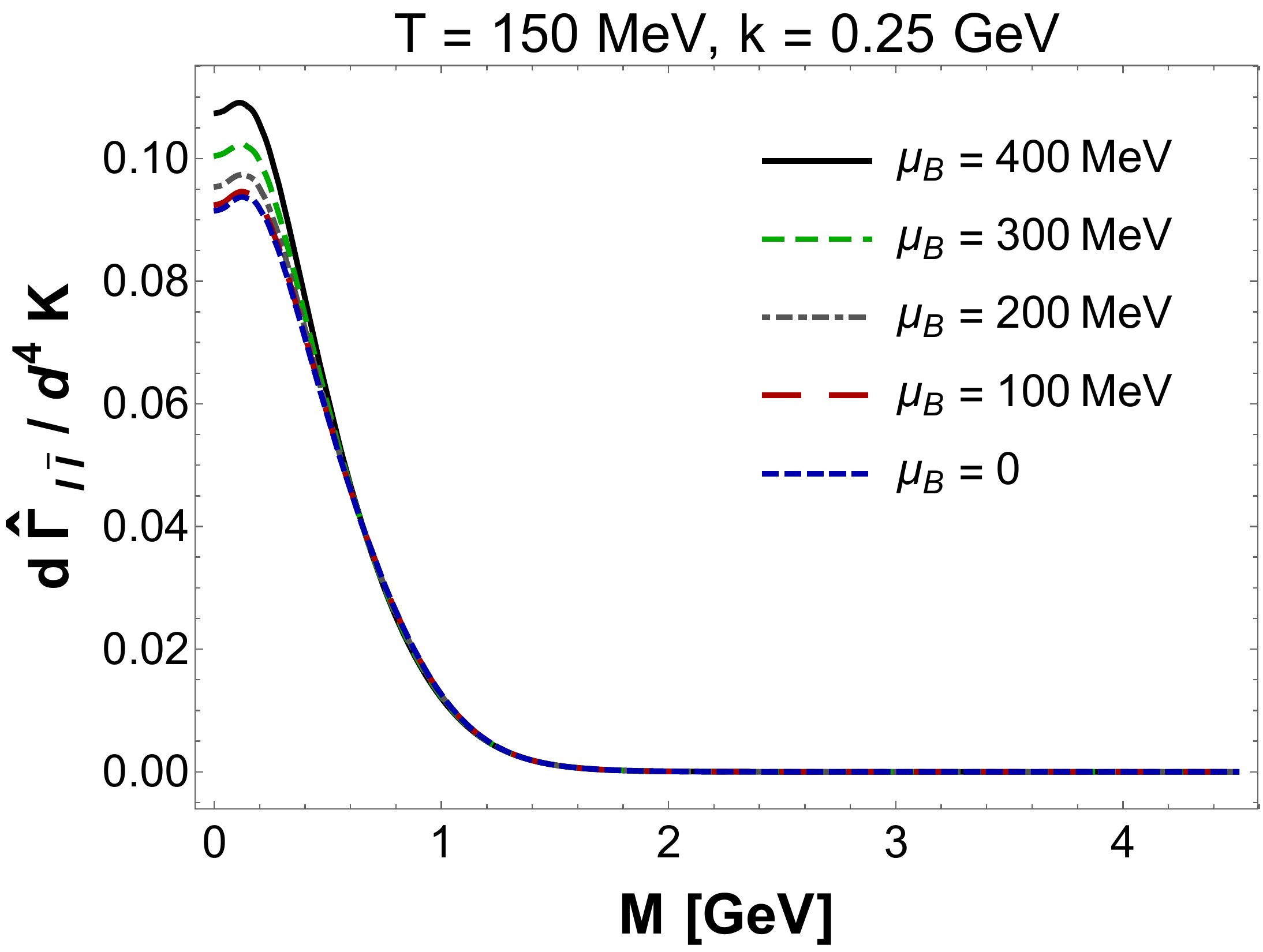}}
\qquad
\subfigure[]{\includegraphics[width=0.46\linewidth]{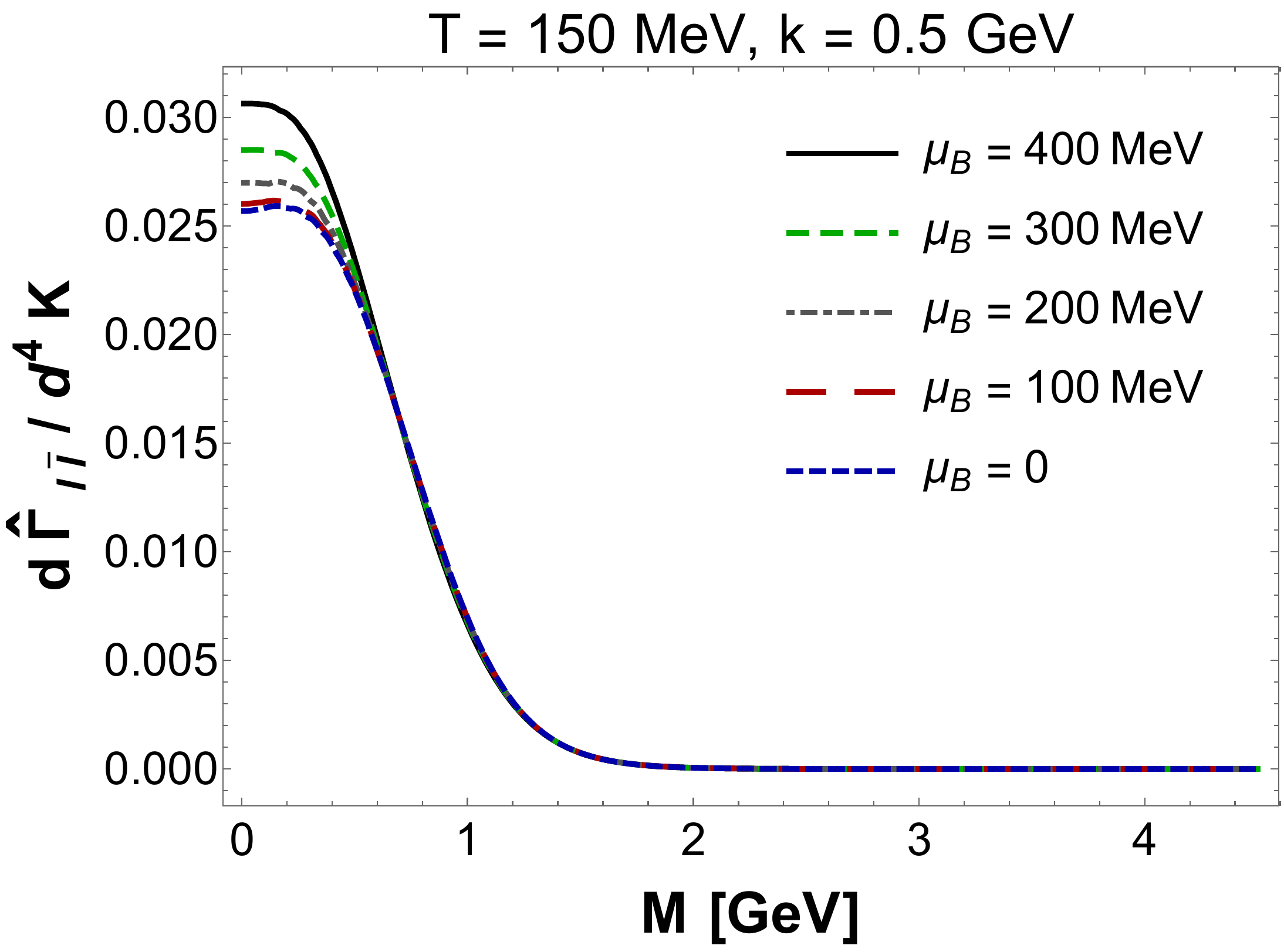}}
\qquad
\subfigure[]{\includegraphics[width=0.46\linewidth]{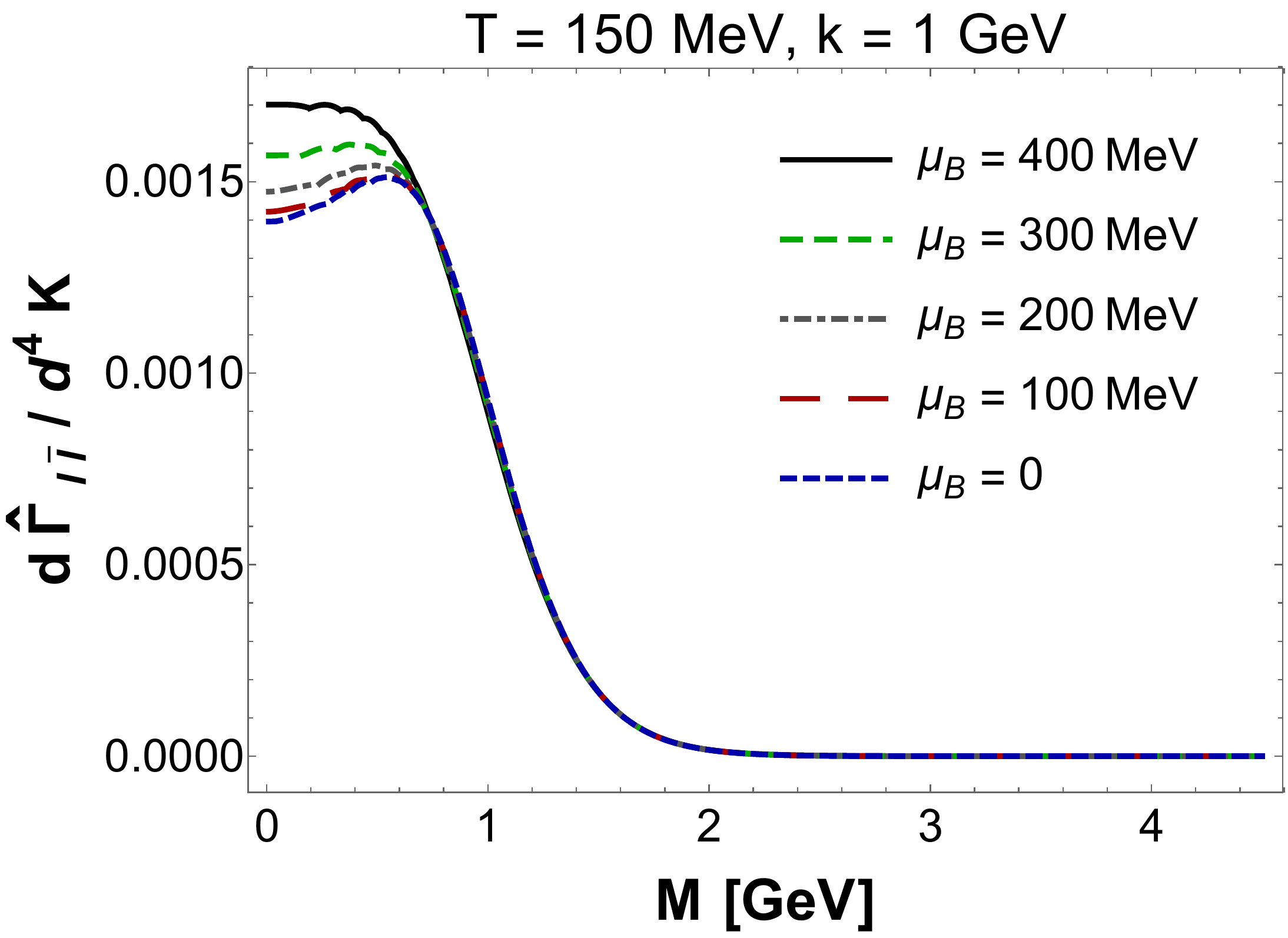}}
\caption{(Color online) Normalized thermal dilepton production rate as a function of the invariant mass $M$ of the dilepton pair for several values of the baryon chemical potential $\mu_B$ at the fixed temperature $T = 150 \, \mathrm{MeV}$ with (a) $k = 0.25 \, \mathrm{GeV}$, (b) $k = 0.5 \, \mathrm{GeV}$, and (c) $k = 1 \, \mathrm{GeV}$.}
\label{fig:thermaldilepton2}
\end{figure}

In Fig. \ref{fig:thermaldilepton}, we show the EMD holographic predictions for the normalized thermal dilepton production rate \eqref{eq:dileptonproduction} as a function of the dilepton invariant mass $M$ for several choices of the temperature $T$ and momentum $k$, at both zero and nonzero baryon chemical potential $\mu_B$. We note that, as also observed for the thermal photon spectrum, the peak of the thermal dilepton spectrum is enhanced as one increases the temperature. Varying $\mu_B$ has essentially no effect on the thermal dilepton production for $M \gtrsim 1 \, \mathrm{GeV}$; however, as one can see in Fig. \ref{fig:thermaldilepton2}, increasing $\mu_B$ enhances the dilepton production rate for $M \lesssim 1 \, \mathrm{GeV}$. From Fig.'s \ref{fig:thermaldilepton} and \ref{fig:thermaldilepton2}, one also notes that increasing the momentum $k$ dampens the thermal dilepton production rate.

\section{Conclusions}
\label{sec:conclusion}

In the present work, we made use of a bottom-up phenomenological Einstein-Maxwell-Dilaton holographic model which is in good quantitative agreement with recent lattice data for the thermodynamics of $(2+1)$-flavor QCD for values of the baryon chemical potential $\mu_B$ up to 400 MeV, corresponding to the maximum value of $\mu_B$ reached in the current beam energy scan at RHIC, and calculate many electric charge transport observables and the thermal spectra of photons and dileptons in a hot and baryon rich strongly coupled QGP.

First, we obtained holographic predictions for the temperature $T$ and baryon chemical potential $\mu_B$ dependence of the electric charge susceptibility, which was found to increase with increasing $\mu_B$. Next, we calculated the DC and AC electric conductivities and also the electric charge diffusion constant as functions of $T$ and $\mu_B$. Around the crossover transition, we found reasonable agreement between our results for the DC electric conductivity and the electric charge diffusion at $\mu_B=0$ and the latest lattice data available \cite{Aarts:2014nba} for these physical observables. We also found that an increase in $T$ and/or $\mu_B$ causes a damping in the oscillations observed in the AC electric conductivity as a function of frequency. Moreover, we obtained that the diffusion of electric charge is suppressed in a baryon rich QGP as one move toward increasing values of the baryon density.

The main results of the present work concern the holographic calculation of the thermal photon and dilepton production rates at both zero and nonzero $\mu_B$. We found that increasing $T$ and/or $\mu_B$ enhances the peak present in both spectra, albeit they are less sensitive to variations in $\mu_B$ than to variations in $T$. In particular, for values of the invariant dilepton mass $M \gtrsim 1 \, \mathrm{GeV}$ the dilepton spectrum is insensitive to variations in $\mu_B$, although for $M \lesssim 1 \, \mathrm{GeV}$ the thermal dilepton production rate increases with increasing values of the baryon chemical potential. We also observe a suppression in the thermal production of dileptons as one increases their momentum. Finally, the fact that our model results for the electric conductivity and the electric diffusion constant at $\mu_B=0$ are in reasonable agreement with the latest lattice data \cite{Aarts:2014nba} for these observables around the crossover transition, gives a constraint on the order of magnitude of the thermal photon and dilepton spectra in a strongly coupled QGP, which according to the present holographic results is typically one order of magnitude lower than perturbative estimates available in the literature.

\acknowledgments

We thank J. Noronha, H. Nastase and R. Critelli for their insightful comments and discussions. S.I.F. was supported by the S\~{a}o Paulo Research Foundation - FAPESP and Coordena\c{c}\~ao de Aperfei\c{c}oamento de Pessoal de N\'{i}vel Superior (CAPES) under FAPESP grant number 2015/00240-7. R.R. acknowledges financial support by FAPESP under FAPESP grant number 2013/04036-0.

\FloatBarrier

\end{document}